%% file: FSQ-16-006_temp.tex
\pdfoutput=1

\documentclass[11pt,twoside,a4paper,cmspaper,final,collab]{cms-tdr}
\def\svnVersion{b9b35d2}\def\svnDate{2020/09/04}\def\cmsCernNoTag{CERN-EP-2020-005}\def\cmsCernDate{\today}\def\cmsMessage{'Published in the European Physical Journal C as \href{http://dx.doi.org/10.1140/epjc/s10052-020-8166-5}{\doi{10.1140/epjc/s10052-020-8166-5}.}'}

\begin{document}\cmsNoteHeader{FSQ-16-006}

\hyphenation{had-ron-i-za-tion}
\hyphenation{cal-or-i-me-ter}
\hyphenation{de-vices}
\newlength\cmsFigWidth
\ifthenelse{\boolean{cms@external}}{\setlength\cmsFigWidth{0.85\columnwidth}}{\setlength\cmsFigWidth{0.4\textwidth}}
\ifthenelse{\boolean{cms@external}}{\providecommand{\cmsLeft}{top\xspace}}{\providecommand{\cmsLeft}{left\xspace}}
\ifthenelse{\boolean{cms@external}}{\providecommand{\cmsRight}{bottom\xspace}}{\providecommand{\cmsRight}{right\xspace}}
\newlength\cmsTabSkip\setlength{\cmsTabSkip}{1ex}
\providecommand{\mub}{\ensuremath{\,\mu\text{b}}\xspace}
\providecommand{\systlum}{\ensuremath{\,\text{(syst+lumi)}}\xspace}
\DeclareRobustCommand{\PGrz}{{\HepParticle{\PGr}{}{0}{}{}{}}\xspace}
\DeclareRobustCommand{\PGccDz}{{\HepParticle{\PGc}{c0}{}}\Xspace}
\DeclareRobustCommand{\PGccDo}{{\HepParticle{\PGc}{c1}{}}\Xspace}
\DeclareRobustCommand{\PGccDt}{{\HepParticle{\PGc}{c2}{}}\Xspace}
\DeclareRobustCommand{\PGcbDz}{{\HepParticle{\PGc}{b0}{}}\Xspace}
\DeclareRobustCommand{\PGcbDo}{{\HepParticle{\PGc}{b1}{}}\Xspace}
\DeclareRobustCommand{\PGcbDt}{{\HepParticle{\PGc}{b2}{}}\Xspace}

\cmsNoteHeader{FSQ-16-006}
\title{Study of central exclusive $\PGpp\PGpm$ production in proton-proton collisions at $\sqrt{s} = 5.02$ and 13\TeV}

\date{\today}

\abstract{
Central exclusive and semiexclusive production of $\PGpp\PGpm$ pairs is measured with the CMS detector in proton-proton collisions at the LHC at center-of-mass energies of 5.02 and 13\TeV. The theoretical description of these nonperturbative processes, which have not yet been measured in detail at the LHC, poses a significant challenge to models. The two pions are measured and identified in the CMS silicon tracker based on specific energy loss, whereas the absence of other particles is ensured by calorimeter information. The total and differential cross sections of exclusive and semiexclusive central $\PGpp\PGpm$ production are measured as functions of invariant mass, transverse momentum, and rapidity of the $\PGpp\PGpm$ system in the fiducial region defined as transverse momentum $\pt(\PGp) > 0.2\GeV$ and pseudorapidity $\abs{\eta(\PGp)} < 2.4$. The production cross sections for the four resonant channels \PfDzP{500}, $\PGrzP{770}$, \PfDzP{980}, and \PfDtP{1270} are extracted using a simple model. These results represent the first measurement of this process at the LHC collision energies of 5.02 and 13\TeV.}

\hypersetup{
pdfauthor={CMS Collaboration},
pdftitle={Study of central exclusive pi+pi- production in proton-proton collisions at sqrt(s) = 5.02 and 13 TeV},
pdfsubject={CMS},
pdfkeywords={CMS, physics, diffraction, exclusive processes}}

\maketitle

\section{Introduction}
The central exclusive production (CEP) process has been studied for a long time from both theoretical \cite{article:lebiedowicz,Lebiedowicz:2016ioh,Lebiedowicz:2014bea,article:bolz,Harland-Lang:2013dia,Harland-Lang:2013qia,Ryutin:2019khx} and experimental \cite{Reyes:1997ei,article:wa102,Barberis:1999ap,Barberis:1999an,Breakstone:1986xd,Breakstone:1990at,Breakstone:1993ku,article:isr-afs,article:cdf,Adloff:2000vm,Chekanov:2009zz} perspectives. In this process, both protons remain intact in the collision and a central system is produced. The process is referred to as exclusive when no particles other than the central system are produced. If one or both protons dissociate into a forward diffractive system, the process is called semiexclusive production. Various central systems can be produced in this process, like $\PGpp\PGpm$, $\PKp\PKm$, and $4\PGp$. In this paper, the $\PGpp\PGpm$ central system is measured. At the CERN LHC energies, the two dominant mechanisms of $\PGpp\PGpm$ production via CEP are \emph{double pomeron exchange} (DPE) and \emph{vector meson photoproduction} (VMP), which are illustrated by the diagrams shown in Fig.~\ref{fig:dpe}. The pomeron ($\mathbb{P}$) is a color singlet object introduced to explain the rise of the inelastic cross section at high collision energies \cite{book:forshaw,book:donnachie}. The quantum numbers of the pomeron constrain the possible central systems in DPE processes, whereas the photon exchange restricts the central system in VMP processes. By functioning as a quantum number filter, the CEP process is suitable to study low-mass resonances, which would be difficult to study otherwise. Furthermore, DPE processes are also suitable to search for glueballs (bound states of gluons without valence quarks), because they provide a gluon-rich environment \cite{article:ochs,Godizov:2018knz}. Another process that could contribute to the same final state is the two-photon fusion $\PGg\PGg\to\PGpp\PGpm$, which is expected to have a much smaller cross section than DPE and VMP processes and gives a negligible contribution \cite{Baltz:2009jk}.

\ifthenelse{\boolean{cms@external}}
{
\begin{figure*}[htb]
\centering
\includegraphics[width=0.9\textwidth]{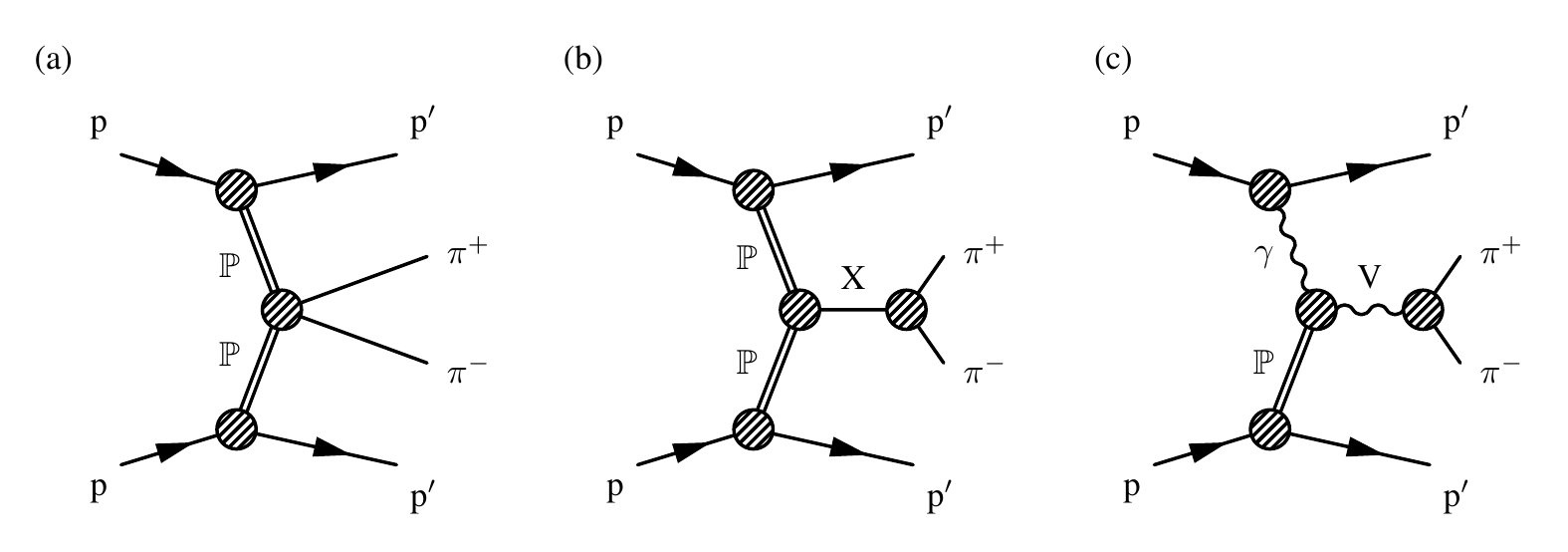}
\caption{Diagrams of the dominant mechanisms for $\PGpp\PGpm$ production via CEP in proton-proton collisions: (a) continuum; (b) resonant double pomeron exchange; and (c) vector meson photoproduction.}
\label{fig:dpe}
\end{figure*}
}
{
\begin{figure*}[!b]
\centering
\includegraphics[width=0.9\textwidth]{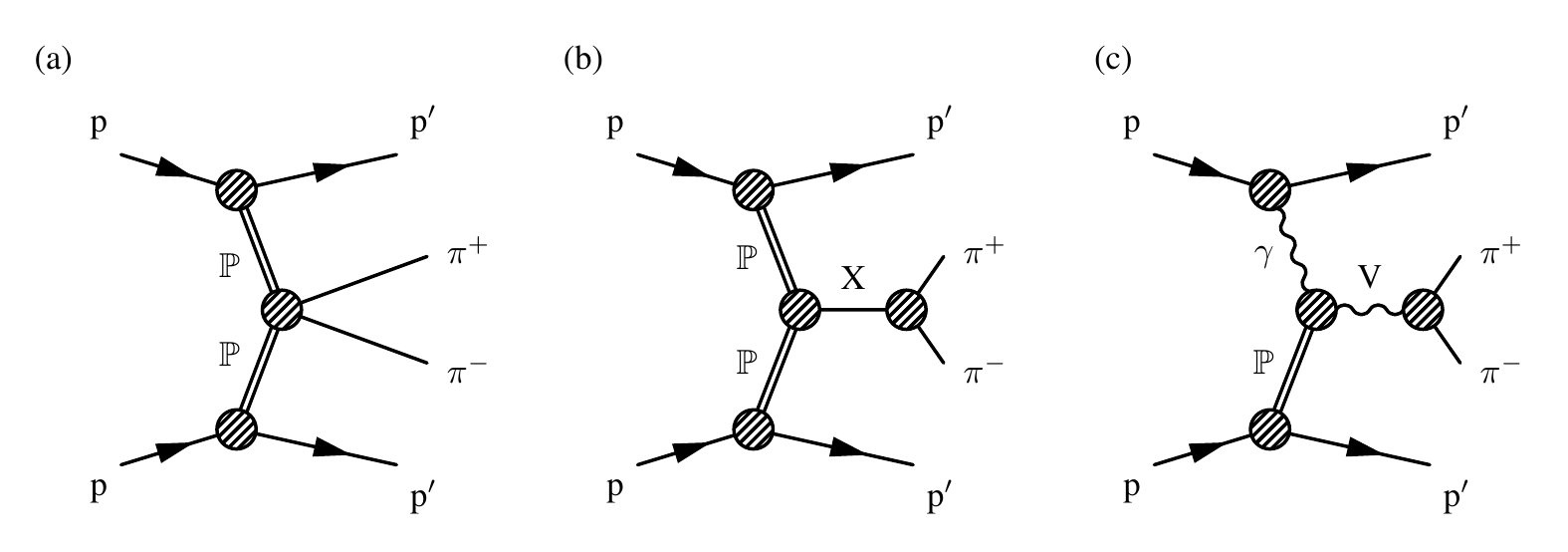}
\caption{Diagrams of the dominant mechanisms for $\PGpp\PGpm$ production via CEP in proton-proton collisions: (a) continuum; (b) resonant double pomeron exchange; and (c) vector meson photoproduction.}
\label{fig:dpe}
\end{figure*}
}

The DPE process of pion pair production has two subcategories: continuum and resonant production. In the case of continuum production, the pion pair is directly produced; thus the pairs have a nonresonant invariant mass spectrum. Resonant production means that an intermediate meson resonance is produced centrally, which manifests itself as a peak in the invariant mass distribution of the pion pair. Since the pomeron is a Regge trajectory running over states with quantum numbers $J^{PC} = \{0^{++},1^{++},2^{++},\dots\}$ and $I^G = 0^+$, the resonance is restricted to have $J^{PC} = \{0^{++}$, $2^{++}$, $4^{++},\dots\}$ and $I^G = 0^+$, where $J$ is the total angular momentum, $I$ is the isospin, $P$ is the parity, $C$ is the charge parity, and $G = C \, (-1)^I$. The known particles \cite{book:pdg} satisfying these criteria are the f$_0$, f$_2$, $\PGccDz$, $\PGccDt$, $\PGcbDz$, and $\PGcbDt$ resonances. 
The cross section for DPE ($\sigma_{\PGpp\PGpm}^\text{DPE}$) can be calculated from the amplitude of continuum ($A_{\PGpp\PGpm}^\text{DPE,C}$) and resonant ($A_{\PGpp\PGpm}^\text{DPE,R}$) production as
\begin{linenomath}
\begin{equation}
\sigma_{\PGpp\PGpm}^\text{DPE} \propto \abs{A_{\PGpp\PGpm}^\text{DPE,C} + A_{\PGpp\PGpm}^\text{DPE,R}}^2.
\end{equation}
\end{linenomath}
Interference terms between the continuum and resonant production channels must be included to describe the observed spectra and to measure the cross sections for resonances.

{\tolerance=1000 In VMP, one of the protons emits a virtual photon, which fluctuates into a quark-antiquark bound state and scatters from the proton via the pomeron exchange. The quantum numbers of the possible resonances are constrained by the quantum numbers of the pomeron and the photon ($J^{PC} = 1^{--}$), leading to mesons with odd spin and the following quantum numbers $J^{PC} = \{1^{--},3^{--},\dots\}$. Resonances satisfying these conditions are $\PGrz$, $\PGo$, $\PGf$, {\JPsi}, $\PGyP{2S}$, and $\PGU$, but only the $\PGrz \to \PGpp\PGpm$ decay has a significant branching fraction, since decays in this channel are strongly suppressed in the case of $\PGf$, {\JPsi}, $\PGyP{2S}$, and $\PGU$ according to the Okubo--Zweig--Iizuka rule \cite{Okubo:1963fa,Zweig:1964jf,Iizuka:1966fk} and in the case of $\PGo$ because of G-parity conservation \cite{OConnell:1995nse}. \par}

This paper presents measurements of exclusive and semiexclusive $\PGpp\PGpm$ total and differential cross sections as functions of invariant mass $m(\PGpp\PGpm)$, transverse momentum $\pt(\PGpp\PGpm)$, and rapidity $y(\PGpp\PGpm)$ of the pion pair, in a fiducial region defined by single pion transverse momentum $\pt(\PGp) > 0.2\GeV$ and single pion pseudorapidity $\abs{\eta(\PGp)} < 2.4$. Because the outgoing protons are not tagged in this measurement, there is a residual contribution from semiexclusive production with all dissociation products outside the $\abs{\eta} > 4.9$ range. In the following, the exclusive and the residual semiexclusive contribution together will be referred to as central exclusive production. The data were recorded by CMS with beam conditions ensuring a small probability of multiple $\Pp\Pp$ collisions in the same bunch crossing (pileup) in August 2015 at a center-of-mass energy of 13\TeV with luminosity 258\mubinv and in November 2015 at 5.02\TeV with a luminosity of 522\mubinv. The average number of $\Pp\Pp$ collisions in a bunch crossing was around 0.3--0.5 for the 5.02\TeV and around 0.5 for the 13\TeV data sets.

\section{The CMS detector}
The central feature of the CMS apparatus is a superconducting solenoid of 6\unit{m} internal diameter. Within the solenoid volume are a tracker, a lead tungstate crystal electromagnetic calorimeter (ECAL), and a brass and scintillator hadron calorimeter (HCAL), each composed of a barrel and two endcap sections, covering the $\abs{\eta} < 3.0$ region. Forward calorimeters extend the $\eta$ coverage provided by the barrel and endcap detectors. Muons are measured in gas-ionization detectors embedded in the steel flux-return yoke outside the solenoid. 

The silicon tracker measures charged particles within the range $\abs{\eta} < 2.5$. It consists of 1440 silicon pixel and 15\,148 silicon strip detector modules and is located in the 3.8\unit{T} solenoid field. Three pixel barrel layers (PXB) are situated at radii of 4.4, 7.3, and 10.2\unit{cm}; PXB also has two pixel endcap disks (PXF). The strip tracker consists of the innermost tracker inner barrel (TIB) and the tracker inner disks (TID), which are surrounded by the tracker outer barrel (TOB). It is completed by endcaps (TEC) on both sides. The barrel part of the strip tracker has a total of 10 layers at radii from 25 to 110\unit{cm}, whereas the endcap of the strip tracker consists of 12 layers. For charged particles with $\pt < 1\GeV$ and $\abs{\eta} < 1.4$, the track resolutions are typically 1--2\% in \pt , and 90--300 and 100--350\mum for the transverse and longitudinal impact parameters, respectively \cite{Chatrchyan:2014fea}. The tracker provides an opportunity to identify charged particles with $0.3 < p < 2\GeV$ based on their specific ionization in the silicon detector elements \cite{Khachatryan:2010pw}.

The ECAL consists of 75\,848 lead tungstate crystals, which provide coverage in $\abs{ \eta } < 1.479 $ in the barrel region and $1.479 < \abs{ \eta } < 3.0$ in the two endcap regions. 

The barrel and endcap sections of the HCAL consist of 36 wedges each and cover the $\abs{\eta} < 3.0$ region. In the region $\abs{ \eta } < 1.74$, the HCAL cells have widths of 0.087 in $\eta$ and 0.087 radians in azimuth ($\phi$). In the $\eta$-$\phi$ plane, and for $\abs{\eta} < 1.48$, the HCAL cells map onto $5{\times}5$ ECAL crystal arrays to form calorimeter towers projecting radially outwards from close to the nominal interaction point. At larger values of $\abs{ \eta }$, the towers are larger and the matching ECAL arrays contain fewer crystals.

The forward hadron (HF) calorimeter uses steel as an absorber and quartz fibers as the sensitive material. The two halves of the HF are located at 11.2\unit{m} from the interaction region, one at each end. Together they provide coverage in the range $3.0 < \abs{\eta} < 5.2$. Each HF calorimeter consists of 432 readout towers, containing long and short quartz fibers running parallel to the beam. The long fibers run the entire depth of the HF calorimeter (165\unit{cm}, or approximately 10 interaction lengths), whereas the short fibers start at a depth of 22\unit{cm} from the front of the detector. By reading out the two sets of fibers separately, it is possible to distinguish showers generated by electrons or photons, which deposit a large fraction of their energy in the long-fiber calorimeter segment, from those generated by hadrons, which typically produce, on average, nearly equal signals in both calorimeter segments.

The triggers used in this analysis are based on signals from the Beam Pick-up and Timing for eXperiments (BPTX) detectors \cite{Khachatryan:2016bia}. The BPTX devices have a time resolution of less than 0.2\unit{ns}. They are located around the beam pipe at a distance of $\pm 175$\unit{m} from the nominal interaction point, and are designed to provide precise information on the bunch structure and timing of the proton beams.

A more detailed description of the CMS detector, together with a definition of the coordinate system used and the relevant kinematic variables, can be found in Ref.~\cite{Chatrchyan:2008zzk}. 

\section{Monte Carlo simulations}
\label{section:mc}
Two kinds of Monte Carlo (MC) event generators are used in this analysis: inclusive and exclusive generators. The inclusive generators model the inclusive diffractive dissociation \cite{Khachatryan:2015gka} and nondiffractive interactions, and are used to estimate the tracking efficiency, multiple reconstruction and misreconstruction rates. The exclusive generators are used to generate CEP events and to calculate the vertex correction factors. There are no available MC event generators that produce exclusive scalar and tensor resonances via DPE, such as the production of \PfDzP{500}, \PfDzP{980}, and \PfDtP{1270} mesons.

Event samples are generated with various tunes for diffraction and the underlying event:
\begin{itemize}
\item \PYTHIA 8.205 \cite{Sjostrand:2014zea} with CUETP8M1 tune \cite{Khachatryan:2015pea} and MBR model \cite{Ciesielski:2012mc}: \PYTHIA \textsc{8} is an inclusive generator based on the Schuler and Sj\"ostrand model. It is capable of modeling a wide variety of physical processes, such as single diffractive (SD), double diffractive (DD), and central diffractive (CD) dissociation, as well as nondiffractive (ND) production \cite{Khachatryan:2015gka}. The SD, DD, and ND events are generated with the CUETP8M1 tune. The Minimum Bias Rockefeller (MBR) model of \PYTHIA is based on the renormalized pomeron flux model and it is capable of generating SD, DD, ND and CD events.
\item \textsc{epos} \cite{Werner:2005jf} with its LHC tune \cite{Pierog:2013ria}: This inclusive generator is based on the Regge--Gribov phenomenology \cite{Drescher:2000ha}, and it models SD, DD, CD, and ND processes.
\item \textsc{starlight} \cite{PhysRevLett.92.142003}: This event generator models photon-photon and photon-pomeron interactions in $\Pp\Pp$ and heavy ion collisions. The production of $\PGrz$ mesons and their successive decay into two pions through the VMP process is simulated by \textsc{starlight}. For background studies, $\PGo$ mesons are also generated with \textsc{starlight} and their decay simulated by \PYTHIA to the $\PGpp\PGpm\PGpz$ final state.
\item \textsc{dime mc} 1.06 \cite{Harland-Lang:2013dia}: The \textsc{dime mc} software describes continuum $\PGpp\PGpm$ production through DPE. The generator uses a phenomenological model based on Regge theory. Events are generated with the Orear-type off-shell meson form factors with parameters $a_{\text{or}} = 0.71\GeV^{-1}$ and $b_{\text{or}} = 0.91\GeV^{-1}$ \cite{Harland-Lang:2013dia}. Furthermore, two additional MC samples are generated with an exponential form factor with $b_{\text{exp}} = 0.45$ \cite{Harland-Lang:2013dia} and $1\GeV^{-2}$ \cite{article:lebiedowicz} to study the systematic uncertainty in the measured resonance cross sections arising from uncertainties in the \textsc{dime mc} parametrization.
\end{itemize}

All of the generated events are processed by a detailed \GEANTfour simulation \cite{Agostinelli:2002hh} of the CMS detector.

\section{Event selection}
The following triggers were employed:
\begin{itemize}
\item Zero bias: zero-bias events are selected by using either the BPTX detectors (13\TeV data) or the LHC clock signal and the known LHC bunch structure (5.02\TeV data). Both methods provided zero-bias events.
\item BPTX XOR: Here XOR stands for the exclusive OR logic, where only one BPTX is fired, corresponding to an incoming proton bunch from only one direction. This trigger was used in both 5.02 and 13\TeV data sets.
\item No-BPTX: There is no signal in the BPTX detectors, which means there are no incoming proton bunches. This trigger was used in both 5.02 and 13\TeV data sets.
\end{itemize}

The present analysis uses events acquired with the zero bias trigger.
The BPTX XOR and No-BPTX triggers select events with no interacting bunches, which are used to estimate the electronic noise of calorimeters and possible collisions between beam particles and residual gas molecules in the CMS beampipe (beam-gas background). The contribution from beam-gas collisions is negligible because there is no difference in the measured calorimeter tower energy distributions for the BPTX XOR and No-BPTX triggered events.

In the offline selections, it is required that the event has exactly two tracks, both of which satisfy $\chi^2/\text{ndf} < 2$ (where the $\chi^2$ value is calculated based on the fitted trajectory and the measured tracker hits, and ndf is the number of degrees of freedom), $\pt > 0.2\GeV$, and $\abs{\eta} < 2.4$ to ensure high track reconstruction efficiency. Only events with oppositely charged (opposite-sign, OS) tracks are selected for analysis, whereas events with same-sign (SS) tracks are used in the background estimation.

Events with a single collision are selected by requiring the two tracks form a single reconstructed vertex subject to the constraint that
\begin{linenomath}
\begin{equation}
\abs{z_1 - z_2} < 3 \sqrt{\sigma_1^2 + \sigma_2^2},
\end{equation}
\end{linenomath}
where $z_1$ and $z_2$ are the $z$ coordinates of the closest approach of the reconstructed tracks to the beamline, and $\sigma_1$ and $\sigma_2$ are their corresponding uncertainties.

To select exclusive events, all calorimeter towers not matching the trajectories of the two tracks must have energy deposits below a threshold, which is defined in Table \ref{tab:threshold}. A tower is matched to a track if the intersection of the extrapolated trajectory with the calorimeter surface is within three standard deviations in $\eta$ and $\phi$ from the center of the tower. The threshold values are chosen to have a maximum 1\% rejection of signal events resulting from the electronic noise of the calorimeters. Non-exclusive events might be also selected because of the lack of coverage in the eta gap between the HF and central calorimeters; these events are also taken into account in the background estimation presented later in this paper.

Using all of the above listed event selection criteria, a total of 48\,961 events were selected from the 5.02\TeV and 20\,980 from the 13\TeV dataset. 

\begin{table}[th]
\centering
\topcaption{The value of calorimeter thresholds for different calorimeter constituents, used in the selection of exclusive events.}
\begin{tabular}{lcr}
\hline
\multicolumn{1}{c}{Calorimeter} & \multicolumn{1}{c}{Threshold [\GeVns{}]} & $\eta$ coverage \\
\hline
ECAL barrel & 0.6 & $\abs{\eta} < 1.5$ \\
ECAL endcap & 3.3 & $1.5 < \abs{\eta} < 3.0$ \\
HCAL barrel & 2.0 & $\abs{\eta} < 1.3$ \\
HCAL endcap & 3.8 & $1.3 < \abs{\eta} < 3.0$\\
HF  		& 4.0 & $3.15 < \abs{\eta} < 5.2$ \\
\hline
\end{tabular}
\label{tab:threshold}
\end{table}

\section{Data analysis}

\subsection{Particle identification}

\begin{figure*}[!bht]
\centering
\includegraphics[width=0.49\textwidth]{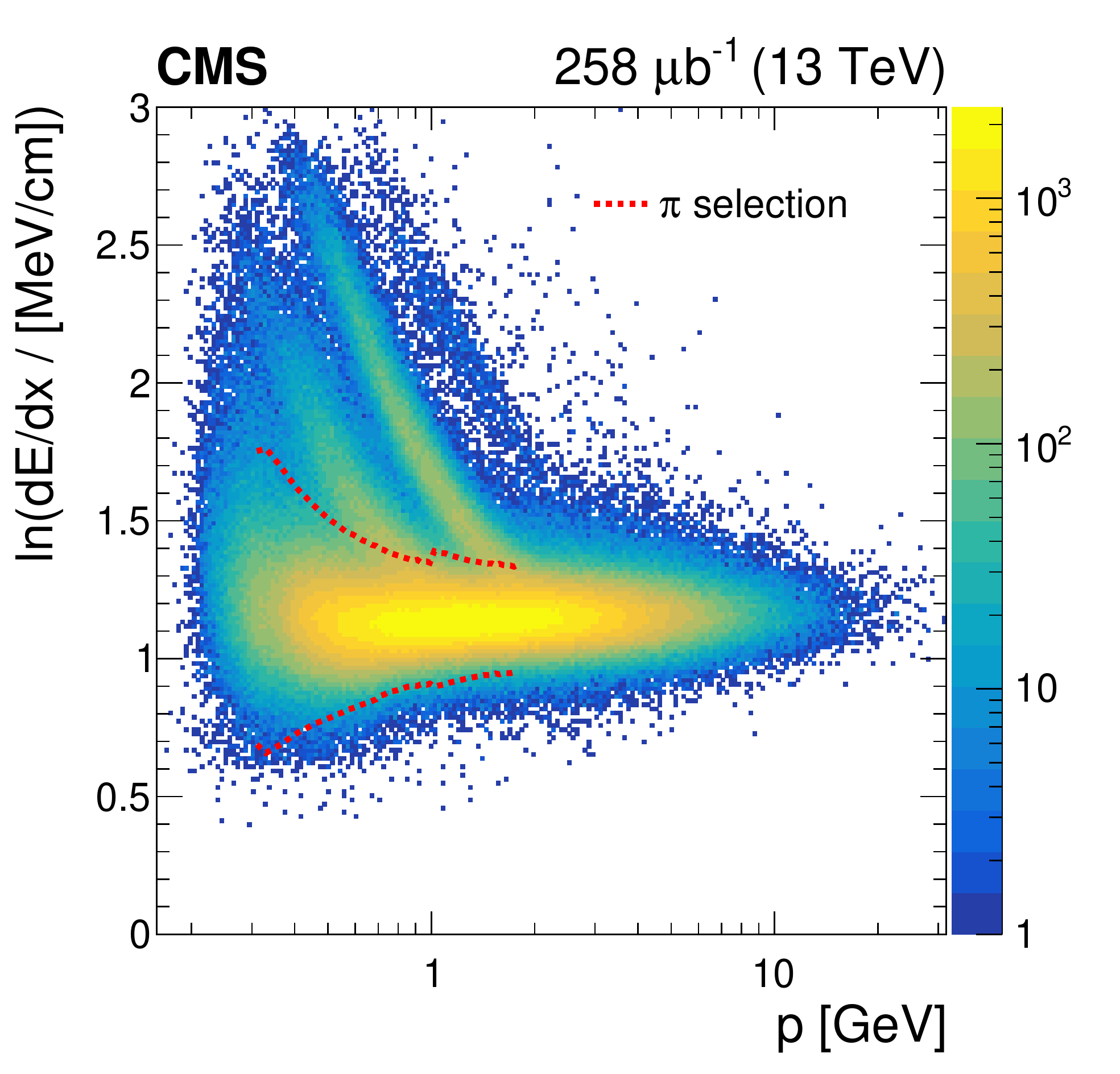} 
\includegraphics[width=0.49\textwidth]{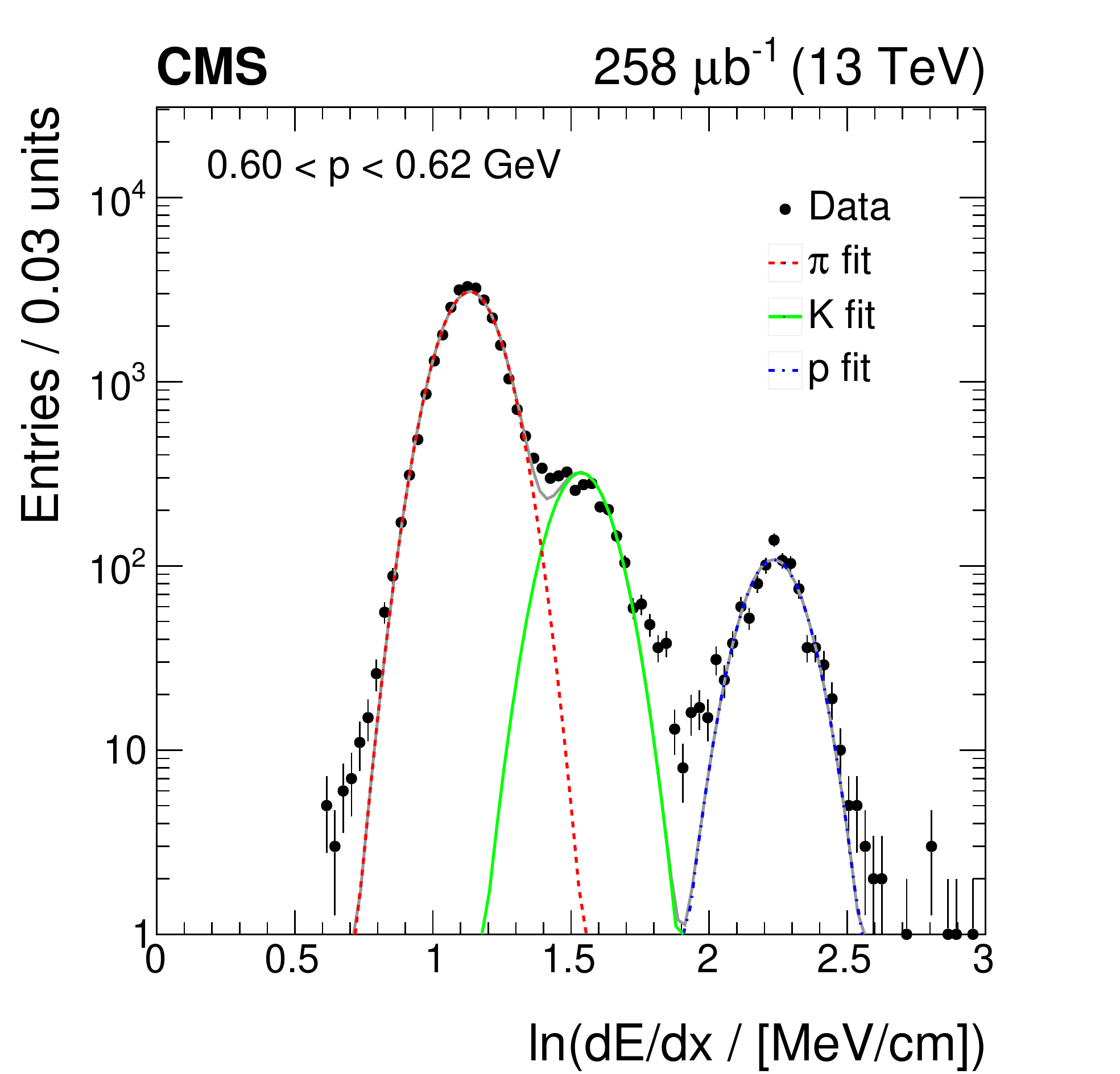} 
\caption{Left: The distribution of the logarithm of the mean energy loss and absolute value of the momentum of tracks from low-multiplicity ($N_\text{track} \leq 4$) events collected at $\sqrt{s} = 13\TeV$. The $\PGp$-selection region is shown in the 0.3--2\GeV range. All tracks outside this momentum range are identified as pions. Right: The fit of energy loss distributions in a given momentum bin with the sum of three Gaussian curves. Plots are similar for the 5.02\TeV data.}
\label{fig:dEdx}
\end{figure*}

Particle identification is used to select pion pairs by the mean energy loss ($\rd E/\rd x$) of particles in the silicon tracking detectors. The $\rd E/\rd x$ values shown in the left panel of Fig.~\ref{fig:dEdx} are calculated by a second-order harmonic mean using only the strip detectors \cite{Quertenmont:2011zz}:
\begin{linenomath}
\begin{equation}
\left\langle \frac{\rd E}{\rd x} \right\rangle = \left( \frac{1}{N} \sum_{i=1}^N (\Delta E / \Delta x)_i^{-2} \right)^{-\frac{1}{2}},
\end{equation}
\end{linenomath}
where $N$ is the number of energy loss measurements, $\Delta E / \Delta x$ is a single energy loss measurement per path length in one tracker module, and the sum runs over the strip detectors carrying energy loss measurements. The $-2$ exponent in this formula suppresses high $\Delta E / \Delta x$ values arising from the highly asymmetric $\Delta E / \Delta x$ Landau distribution, thus avoiding a bias in the estimate of the average $\rd E / \rd x$ of the track.

The track classification is achieved by fitting the mean energy loss distributions of tracks from low multiplicity ($N_\text{track} \leq 4$) events with a sum of three Gaussian functions corresponding to pions, kaons, and protons. An example for such a fit is shown in the right panel of Fig.~\ref{fig:dEdx}. In the 0.3--2\GeV momentum range pions are selected from the $\pm 3$ standard deviation region of the corresponding Gaussian peak. This region is shown in the left panel of Fig.~\ref{fig:dEdx}. Tracks that have $p < 0.3$ or $p > 2\GeV$ are assumed to be pions. The contamination from kaons and protons is estimated using the data-driven approach described in Section \ref{sec:bg}.

\subsection{Corrections}
\label{sec:corr}

Each event is weighted by several correction factors to compensate for the detector and reconstruction effects. The multiplying factor is the product of four independent corrections: tracking, multiple reconstruction, vertex, and pileup correction.

A tracking correction is used to correct for track reconstruction inefficiencies:
\begin{linenomath}
\begin{equation}
C_\text{tr} = \frac{1}{\varepsilon_{\text{tr},1}} \, \frac{1}{\varepsilon_{\text{tr},2}},
\end{equation}
\end{linenomath}
where $\varepsilon_{\text{tr},1}$ ($\varepsilon_{\text{tr},2}$) is the tracking efficiency in the region where the first (second) particle is reconstructed.
A single charged particle may lead to two reconstructed tracks, such as spiralling tracks near $\eta \approx 0$ or split tracks in the overlap region of the tracker barrel and endcap.
This effect is corrected  using $\varepsilon_\text{mrec}$, which is the probability for this situation to occur. In this case the correction factor takes the form
\begin{linenomath}
\begin{equation}
C_\text{mrec} = \frac{1}{1-\varepsilon_{\text{mrec},1}} \, \frac{1}{1-\varepsilon_{\text{mrec},2}}.
\end{equation}
\end{linenomath}
The values of $\varepsilon_\text{tr}$ and $\varepsilon_\text{mrec}$ are estimated as a function of $\eta$ and \pt using MC simulations. Their dependence on the track $\phi$ and the vertex position $z$-coordinate is integrated over. The simulated events are weighted such that the vertex $z$-coordinate distribution agrees with collision data.

The vertex correction $C_\text{vert}$ accounts for events with an unreconstructed vertex. It is the reciprocal of the vertex efficiency, which is calculated using samples produced by the \textsc{dime mc} and \textsc{starlight} generators. The vertex efficiency has a slight dependence on the invariant mass of the track pair that is included when applying the vertex correction.

Some real CEP events are rejected because of pileup. To account for these lost events, a correction factor $C_\text{pu}$ for the number of selected events can be computed. 
The CEP events are selected from bunch crossings with a single collision, so by assuming that the number of collisions follows a Poisson distribution, one can derive $C_\text{pu}$: 
\begin{linenomath}
\begin{equation}
C_\text{pu} = \frac{N \mu}{N \, \mu \exp{(-\mu)}} = \exp{(\mu)}.
\end{equation}
\end{linenomath}
Here, $\mu$ is the average number of visible inelastic collisions, in a given bunch crossing, $N$ is the total number of analyzed events. The value of $\mu$ depends on the instantaneous luminosity associated with individual bunch crossings, $\mathcal{L}_\text{bunch}$, according to the following expression: 
\begin{linenomath}
\begin{equation}
\mu = \frac{\sigma_{\text{inel,vis}}\mathcal{L}_\text{bunch}}{f},
\end{equation}
\end{linenomath}
where $\sigma_\text{inel,vis}$ is the visible inelastic $\Pp\Pp$ cross section, $f$ is the revolution frequency of protons, and $\mathcal{L}_\text{bunch}$ is the average instantaneous luminosity at the given bunch crossing position for time periods of 23.3\unit{s}. The ratio of $\sigma_\text{inel,vis}$ to $f$ is obtained by fitting the fraction of events with no observed collisions as a function of $\mathcal{L}_\text{bunch}$ with the functional form $A\exp(-b \, \mathcal{L}_\text{bunch})$, where $A$ and $b$ are free parameters of the fit.

The range of correction factors is summarized in Table \ref{tab:correction}.

\begin{table}[bt]
\centering
\topcaption{Correction factors.}
\begin{tabular}{lcc}
\hline
\multicolumn{1}{c}{Type} & \multicolumn{1}{c}{Range} \\
\hline
Tracking & 1.05--1.50 \\
Multiple reconstruction & 1.005--1.040 \\
Vertex & 1.05--1.33 \\
Pileup & 1.3--2.1 \\
\hline
\end{tabular}
\label{tab:correction}
\end{table}

\begin{figure*}[htb]
\centering
\includegraphics[width=0.49\textwidth]{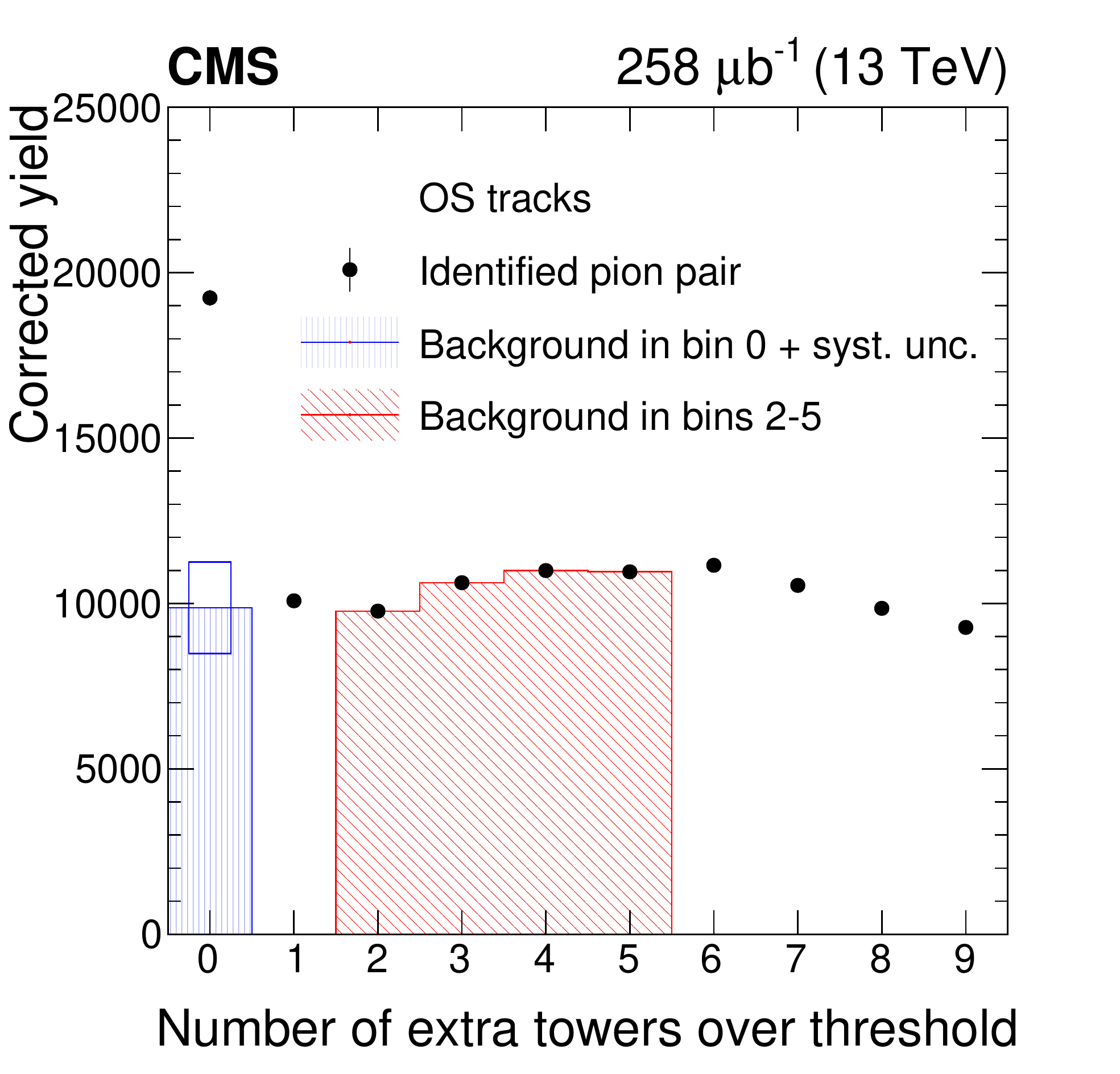} 
\includegraphics[width=0.49\textwidth]{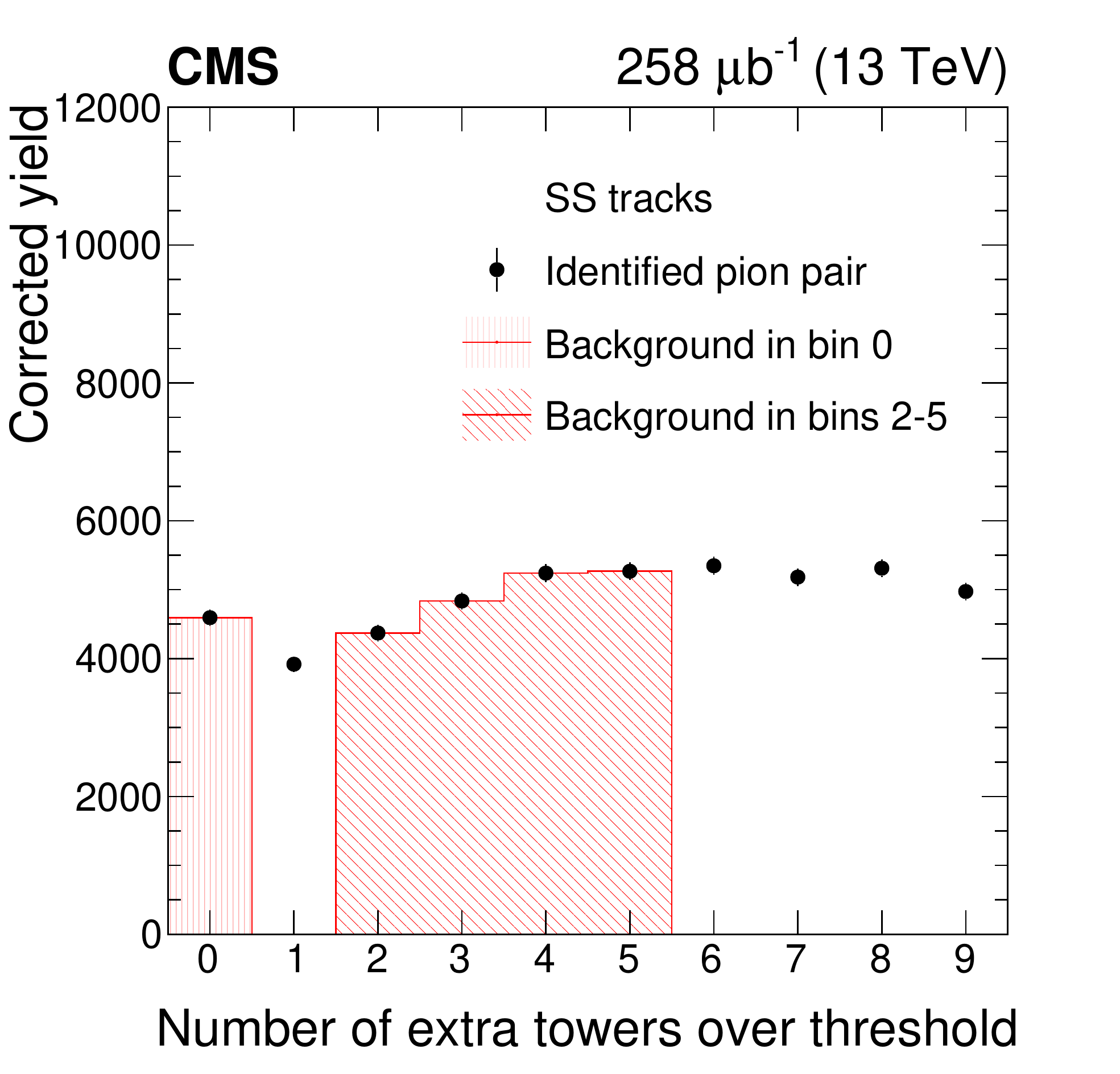} 
\caption{The number of extra calorimeter towers over threshold in events containing an identified pion pair with opposite (left) and same (right) charge. The known contributions, denoted with the red hatched areas, are used to estimate the background in the zero bin of the opposite-sign distribution, which is denoted by the blue hatched area. The error bars correspond to statistical uncertainties, whereas the error rectangle on the background denotes the 14\% systematic uncertainty in the background normalization. Plots are similar for 5.02\TeV data.}
\label{fig:bg}
\end{figure*}

\subsection{Background estimation}
\label{sec:bg}

The main background contributions to $\PGpp\PGpm$ CEP are the multiparticle background and the exclusive $\PKp\PKm$/$\Pp\PAp$ production. The multiparticle background in the selected exclusive sample consists of events with more than two particles created in the interaction, of which only two are observed because the additional particles yield energy deposits below the thresholds, or outside the acceptance. The SD, DD, ND, and CD processes with more than two centrally produced particles belong to this contribution. A method based on control regions is used to estimate this multiparticle background. Control regions are selected in which events have at least two calorimeter towers above threshold, not matched to the two selected pions, with all the other selection criteria satisfied. The distribution of the number of events selected in this way as a function of the number of extra towers with energy above threshold is shown in Fig.~\ref{fig:bg}. The counts in the bins with 2, 3, 4, and 5 towers are used to estimate the background. The normalization factor is calculated using the following assumption:
\begin{linenomath}
\ifthenelse{\boolean{cms@external}}
{
  \begin{multline}
\frac{N_{\text{mpart,SS}}(\text{0 extra towers})}{N_{\text{mpart,SS}}(\text{2--5 extra towers})}  \\
= \frac{N_{\text{mpart,OS}}(\text{0 extra towers})}{N_{\text{mpart,OS}}(\text{2--5 extra towers})},
\label{eq:bg}
\end{multline}}
{
  \begin{equation}
\frac{N_{\text{mpart,SS}}(\text{0 extra towers})}{N_{\text{mpart,SS}}(\text{2--5 extra towers})} = \frac{N_{\text{mpart,OS}}(\text{0 extra towers})}{N_{\text{mpart,OS}}(\text{2--5 extra towers})},
\label{eq:bg}
\end{equation}}
\end{linenomath}
where $N_{\text{mpart,OS/SS}}$ is the number of multiparticle events with two OS or SS tracks. The validity of this assumption is checked by comparing the true and predicted number of background events in inclusive MC samples (Table~\ref{tab:background}). The observed discrepancy reflects the differences between OS and SS events and is included as a systematic uncertainty in the estimate of the total number of multiparticle background events, as discussed in Section \ref{section:syst}. With this formula and the fact that all SS events are multiparticle events because of charge conservation, it is possible to calculate the value of $N_{\text{mpart,OS}}(\text{0 towers})$, which is the number of multiparticle background events. The expected distribution of the multiparticle background is obtained using OS events with 2--5 extra calorimeter towers.

This method does not take into account the background contribution from $\PGo \to \PGpp\PGpm\PGpz$, because this decay cannot be observed in the SS events. This latter contribution is negligible (0.5\%) based on MC simulation results. 

\begin{table}[th]
\centering
\topcaption{Checking the validity of Eq.~(\ref{eq:bg}) by comparing the true and predicted number of background events in inclusive MC samples.}
\begin{tabular}{lr}
\hline
\multicolumn{1}{c}{Event generator} & \multicolumn{1}{c}{Difference in normalization} \\
\hline
\textsc{epos} & $(+11 \pm 4)\%$ \\
\PYTHIA 8 CUETP8M1 & $(-5.5 \pm 3)\%$ \\
\PYTHIA 8 MBR & $(+10 \pm 4)\%$ \\
\hline
\end{tabular}
\label{tab:background}
\end{table}

\begin{figure*}[htb]
\centering
\includegraphics[width=0.49\textwidth]{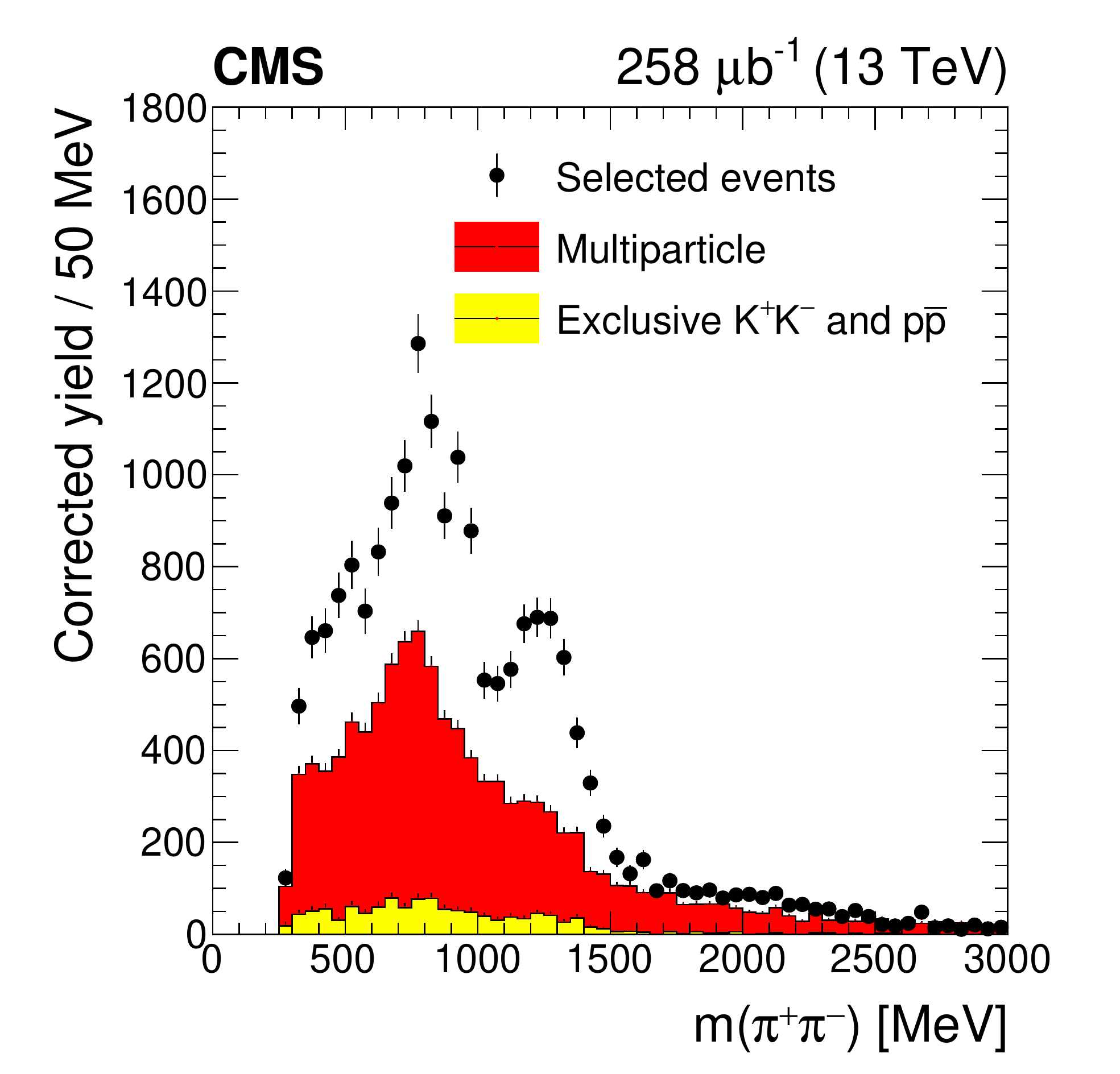} 
\includegraphics[width=0.49\textwidth]{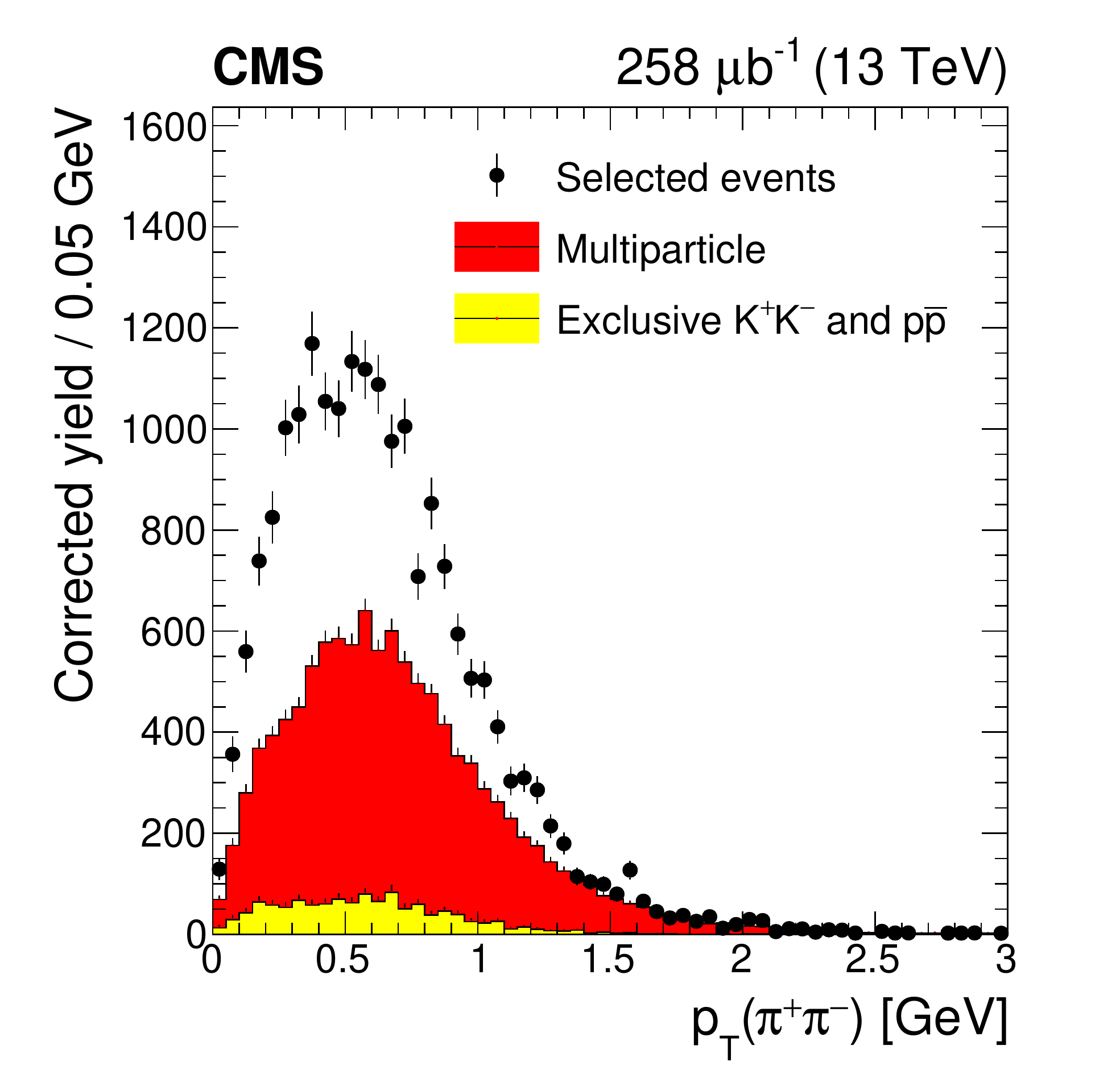} 
\includegraphics[width=0.49\textwidth]{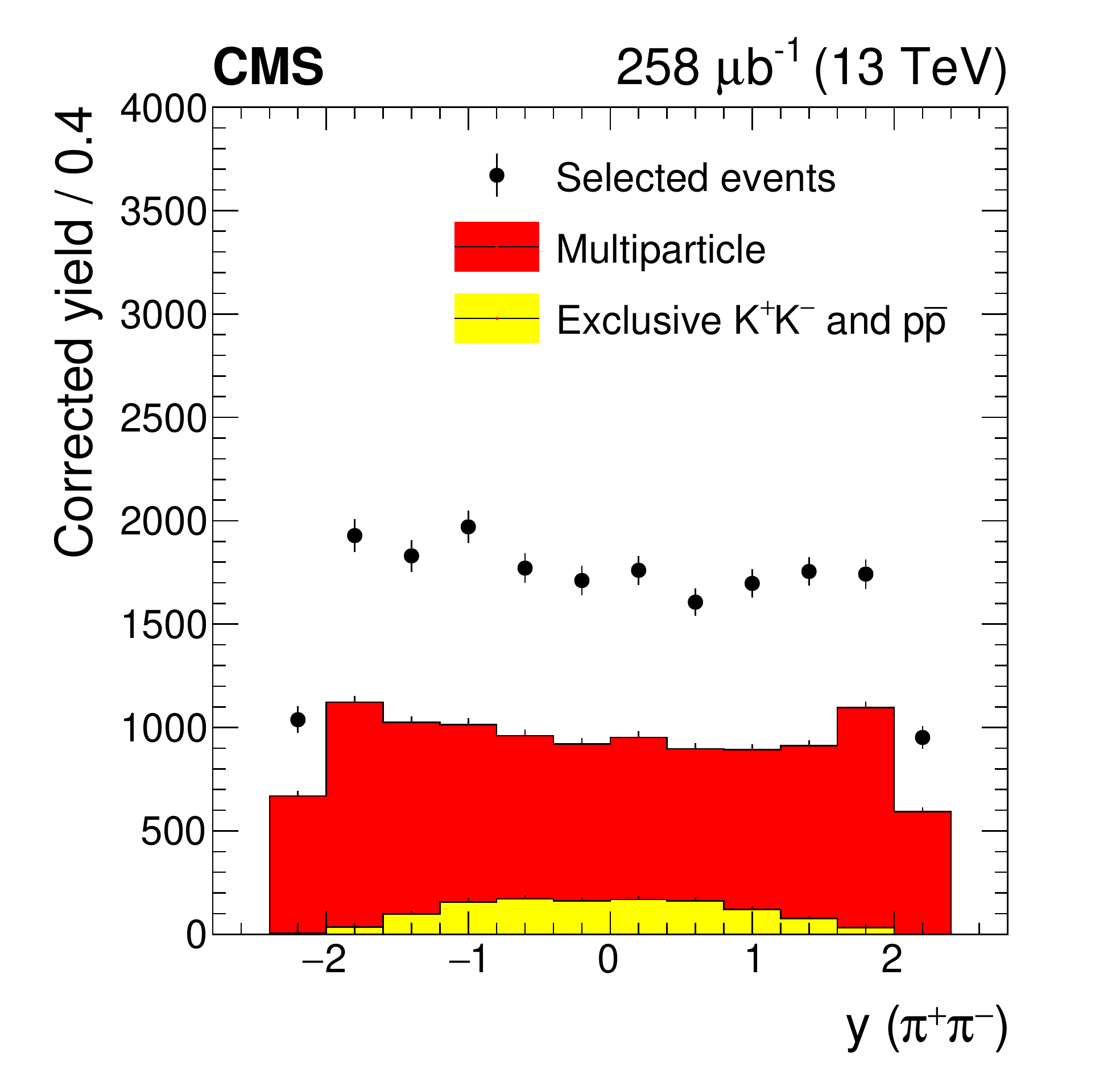} 
\caption{Background distributions as functions of kinematic variables estimated by data-driven methods. The proton dissociation background is not shown here, since it is included via scaling of the final cross section values. The error bars correspond to statistical uncertainties. The results for the 5.02\TeV data set are similar.}
\label{fig:bg_shape}
\end{figure*}

Genuine exclusive $\PKp\PKm$ and $\Pp\PAp$ events, where both particles are misidentified as pions, are included in the previous multiparticle background estimate. To correct for this contribution, the $\PK/\PGp$ ratios are calculated in the exclusive events using tracks with $p < 1\GeV$. Similarly, the $\Pp/\PGp$ ratio is calculated in the same sample in the range  $1 < p < 2\GeV$. The $\PK/\PGp$ and $\Pp/\PGp$ ratios are assumed to be $0.3^{+0.1}_{-0.05}$ in the region $p > 1$ and $p > 2\GeV$, respectively \cite{Chatrchyan:2012qb}. Using this assumption and the measured ratios, the average $\PK/\PGp$ and $\Pp/\PGp$ ratios are then calculated over the entire momentum range of the exclusive sample. These average ratios can then be used to compute the number of $\PKp\PKm$ and $\Pp\PAp$ events under two extreme scenarios. The first scenario assumes that the production of a $\PK$ or a $\Pp$ is always accompanied by the production of its antiparticle, whereas in the second scenario it is assumed that the production of an individual $\PKp$, $\PKm$, $\Pp$, or $\PAp$ is a totally independent process. The final estimate of the exclusive $\PKp\PKm$ and $\Pp\PAp$ background normalization is calculated as the average of the estimates obtained from assuming these two scenarios. According to these calculations, there is an 11\% residual contribution of exclusive $\PKp\PKm$ and $\Pp\PAp$ events in the sample after the multiparticle background subtraction. The background distributions of this contribution are calculated by using two-track OS exclusive events with at least one identified $\PKpm$ (Fig.~\ref{fig:bg_shape}).

The estimated multiparticle and exclusive $\PKp\PKm$/$\Pp\PAp$ background distributions, as functions of the main kinematic variables, are shown in Fig.~\ref{fig:bg_shape}. These two background contributions are subtracted from the measured distributions. The background subtracted spectra are divided by the integrated luminosity to obtain the differential cross sections.

\subsection{Systematic uncertainties}
\label{section:syst}

Systematic uncertainties in the measured cross sections originate from various sources. These include reconstruction effects, particle identification, correction factors, background estimation, and the luminosity estimation. The uncertainty assigned to the tracking efficiency in the case of a single track is 3.9\% \cite{Chatrchyan:2014fea}, which corresponds to 7.8\% uncertainty for two tracks. Furthermore, the uncertainty in the multiple reconstruction rate for a single track is also 3.9\%, which propagates to a maximum of 0.4\% uncertainty in the cross section for two tracks, which is neglected in the analysis. Misreconstructed tracks bias the sample in two ways: either a CEP event is rejected if a third misreconstructed track is found, or an event is identified as CEP with a misreconstructed and a genuine track. This source of systematic uncertainty is estimated to be 1\% for a single track, which is the maximal misreconstruction rate calculated using inclusive MC samples in the kinematic region ($\pt(\PGp) > 0.2\GeV$ and $\abs{\eta(\PGp)} < 2.4$) of the analysis. Since the probability to have two or more misreconstructed tracks in these low-multiplicity events is negligible, the final uncertainty remains 1\%. From the comparison of the \textsc{dime mc} and \textsc{starlight} simulations, the uncertainty of the vertex correction is estimated to be 1\%.

The systematic uncertainty in the pileup correction factor for a single event is calculated from only the systematic uncertainties in the luminosity measurement that do not affect its overall normalization. Indeed, the normalization-related systematic uncertainties are compensated in the exponential fit described in Section \ref{sec:corr}. The uncertainties that do not affect the normalization are estimated to be 1.6\% and 1.5\% for 5.02 \cite{CMS-PAS-LUM-16-001} and 13\TeV \cite{CMS-PAS-LUM-15-001} data, respectively. These values propagate to a 1\% uncertainty in the pileup correction factor for a single event. After adding up all the selected events, the pileup uncertainty becomes smaller than 0.1\%, which is neglected in the following.

The measured signal yield is affected by the uncertainty arising from the two effects associated with calorimeter noise and veto inefficiency caused by the adopted energy thresholds. A genuine CEP event can be erroneously discarded if the calorimeter noise appears above the energy thresholds used in the veto. Conversely a nonCEP event can pass the final selection if the extra particles pass the veto requirements. In the HF, these uncertainties are estimated by varying the calorimeter energy thresholds by $\pm 10\%$ \cite{Chatrchyan:2011wm}. The resulting uncertainty is estimated to be 3\% for both the 5.02 and 13\TeV data sets. Similarly, the ECAL and HCAL thresholds are varied by $\pm 5\%$ \cite{Chatrchyan:2013dga,Sirunyan:2019djp}, which results in a 1\% uncertainty in the corrected yields at both energies.

The systematic uncertainty estimation of the multiparticle background is done by varying the control region used in the background estimation procedure: 1--2, 2--9, and 5--9 extra towers. The estimate of the systematic uncertainty in the multiparticle background normalization is 10\%. Additionally, a 10\% uncertainty is added to this value quadratically, taking into account the deviations shown in Table \ref{tab:background}; thus the final uncertainty in the multiparticle background normalization is 14\%. After subtracting this contribution, this propagates to systematic uncertainties depending on the invariant mass, transverse momentum and rapidity of the pion pair. The multiparticle background estimation uncertainty varies between 10--20\% below $1500\MeV$. Over $1500\MeV$ the uncertainty varies between 20--60\%, because the signal versus background ratio is much smaller. The average uncertainty, used as the systematic uncertainty of the total cross section, is 15\%.

The exclusive $\PKp\PKm$ and $\Pp\PAp$ background uncertainty comes from three sources: (i) multiparticle contamination in the $\rd E/\rd x$ vs.\ momentum distribution that modifies the $\PK/\PGp$ and $\Pp/\PGp$ ratios, (ii) the uncertainty in the $\PK/\PGp$ ratio above 1\GeV, and (iii) the uncertainty in the $\Pp/\PGp$ ratio above 2\GeV. The multiparticle contamination is estimated by checking the difference between two extreme cases: all particle types are produced independently, or the sample is purely exclusive. The results correspond to an uncertainty of 70\% in the normalization of this background contribution at both energies. To account for the uncertainty of $\PK/\PGp$ above 1\GeV and $\Pp/\PGp$ over 2\GeV, the exclusive background normalization is calculated assuming different values (0.25, 0.30, and 0.40 \cite{Chatrchyan:2012qb}) for the $\PK/\PGp$ and $\Pp/\PGp$ ratios in these regions. The uncertainties assigned to these effects are 16 and 4\%, respectively. Thus the total systematic uncertainty of the exclusive $\PKp\PKm$ and $\Pp\PAp$ background normalization is 72\%. After subtracting this background contribution, this propagates to systematic uncertainties, which depend on the invariant mass, transverse momentum, and rapidity of the pion pair. The typical range of this systematic uncertainty contribution is 5--20\%. For the total cross section, this source contributes to an average uncertainty of 6\%. 

\begin{table}[!tbh]
\centering
\topcaption{The sources and average values of systematic uncertainties, used as the systematic uncertainty of the total cross section.}
\begin{tabular}{lr}
\hline
\multicolumn{1}{c}{Source} & \multicolumn{1}{c}{Average value} \\
\hline
Tracking efficiency & 7.8\% \\
Misreconstructed tracks & 1\% \\
Vertex & 1\% \\
HF energy scale & 3\% \\
ECAL and HCAL energy scale & 1\% \\
Multiparticle background & 15\% \\
Exclusive $\PKp\PKm$ and $\Pp\PAp$ background & 6\% \\[\cmsTabSkip]
Total w/o int. luminosity & 18.3\% \\
+ Integrated luminosity   & 2.3\% \\
\hline
\end{tabular}
\label{tab:syst}
\end{table}

All of the systematic uncertainties listed above are the same for the 5.02 and 13\TeV data sets. Additionally, the systematic uncertainty in the integrated luminosity is 2.3\% \cite{CMS-PAS-LUM-16-001,CMS-PAS-LUM-15-001}. The average values of the systematic uncertainties are summarized in Table \ref{tab:syst}. The total systematic uncertainty is obtained by adding the individual contributions in quadrature. All systematic uncertainty contributions are considered fully correlated across invariant mass bins.

\section{Results}

\begin{figure*}[p!]
\centering
\includegraphics[width=0.45\textwidth]{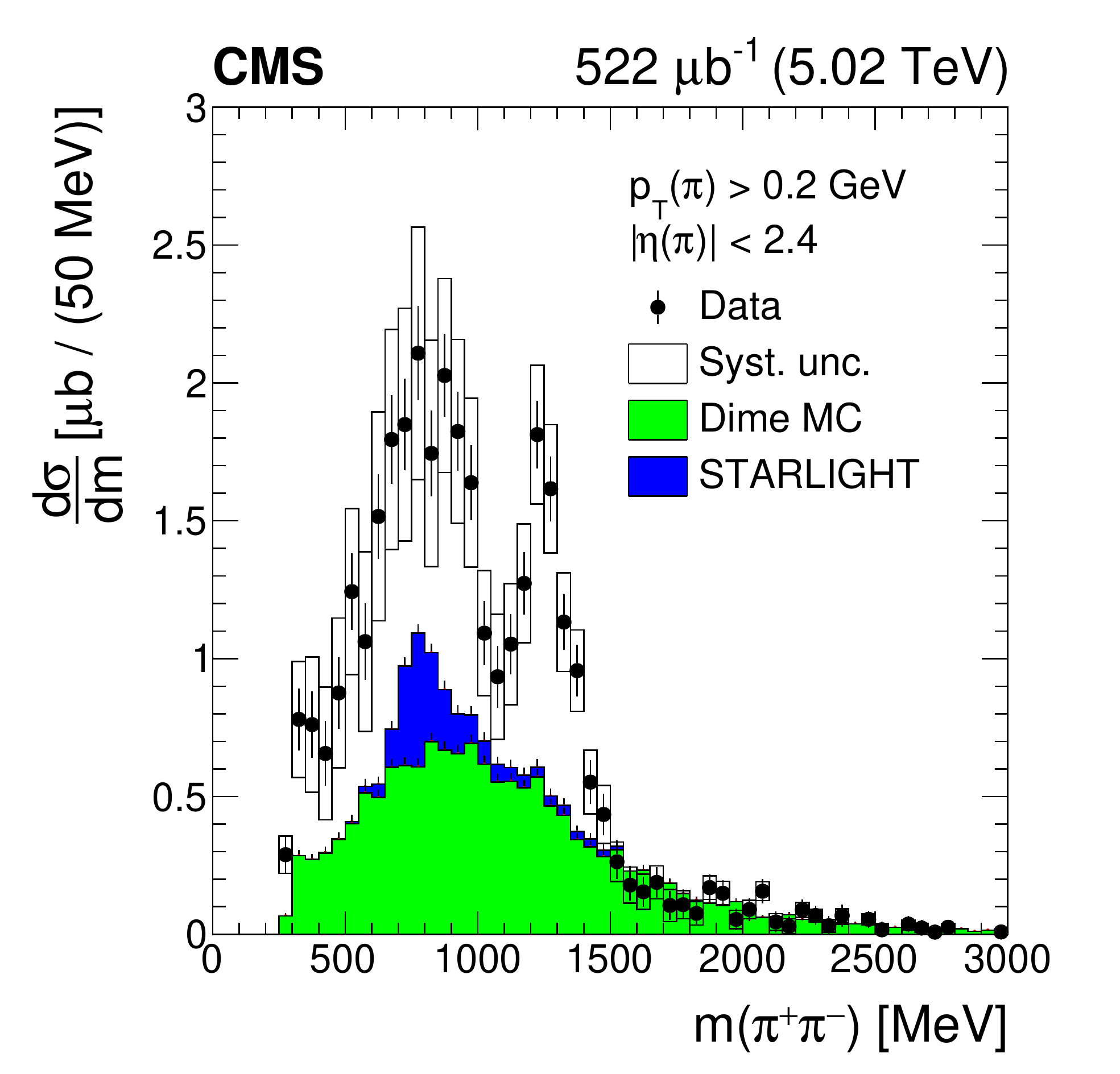}
\includegraphics[width=0.45\textwidth]{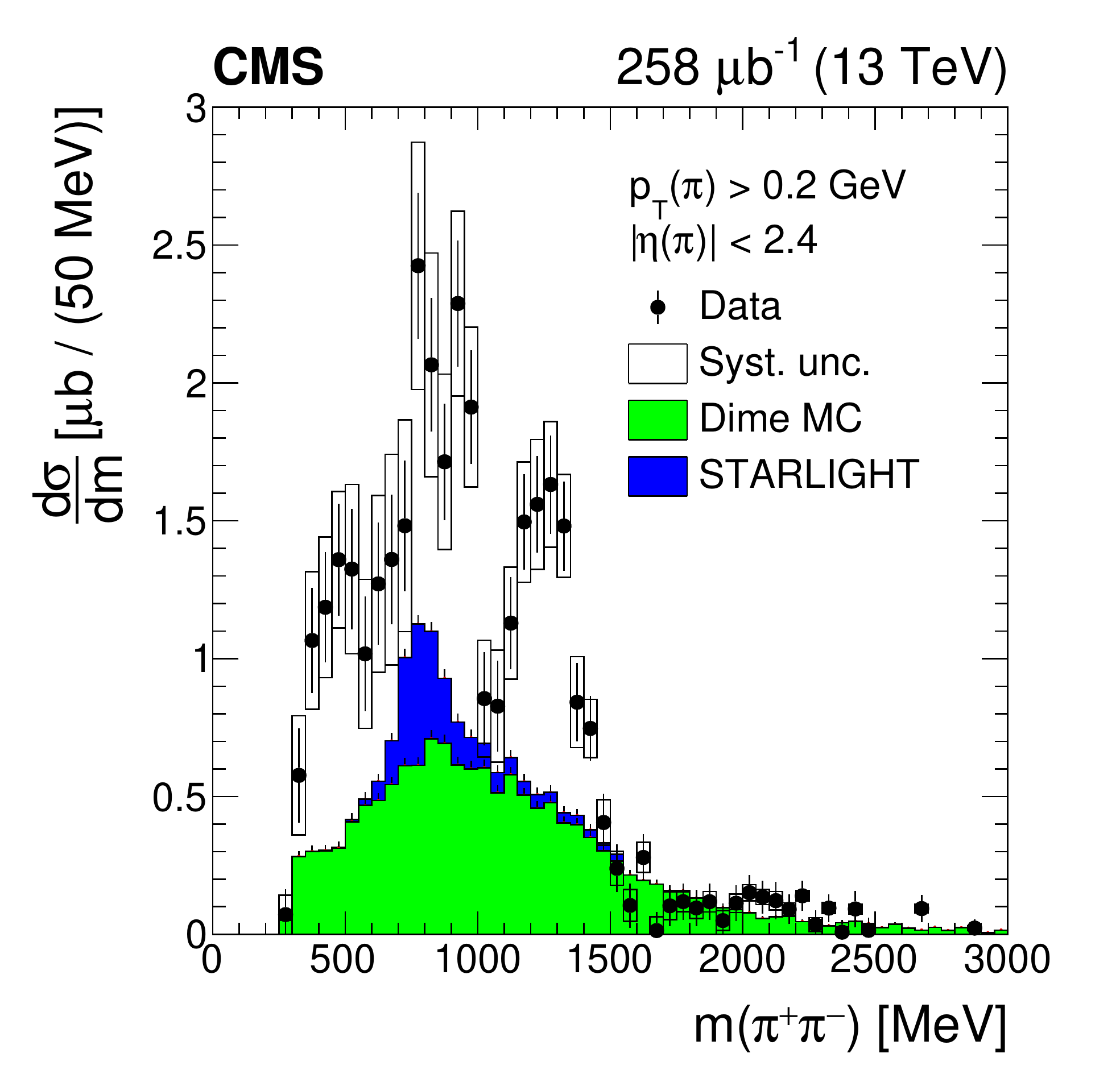}
\includegraphics[width=0.45\textwidth]{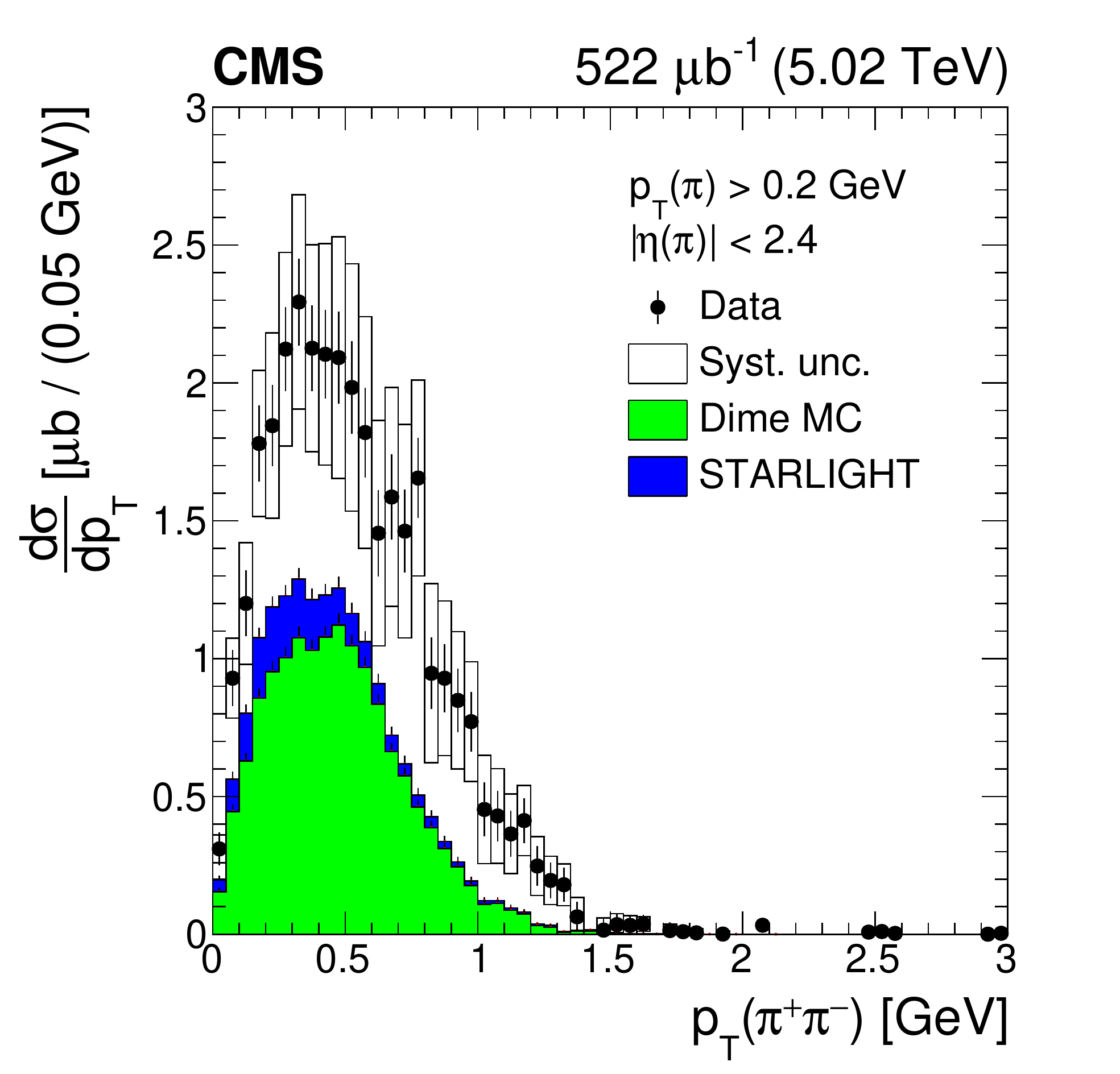}
\includegraphics[width=0.45\textwidth]{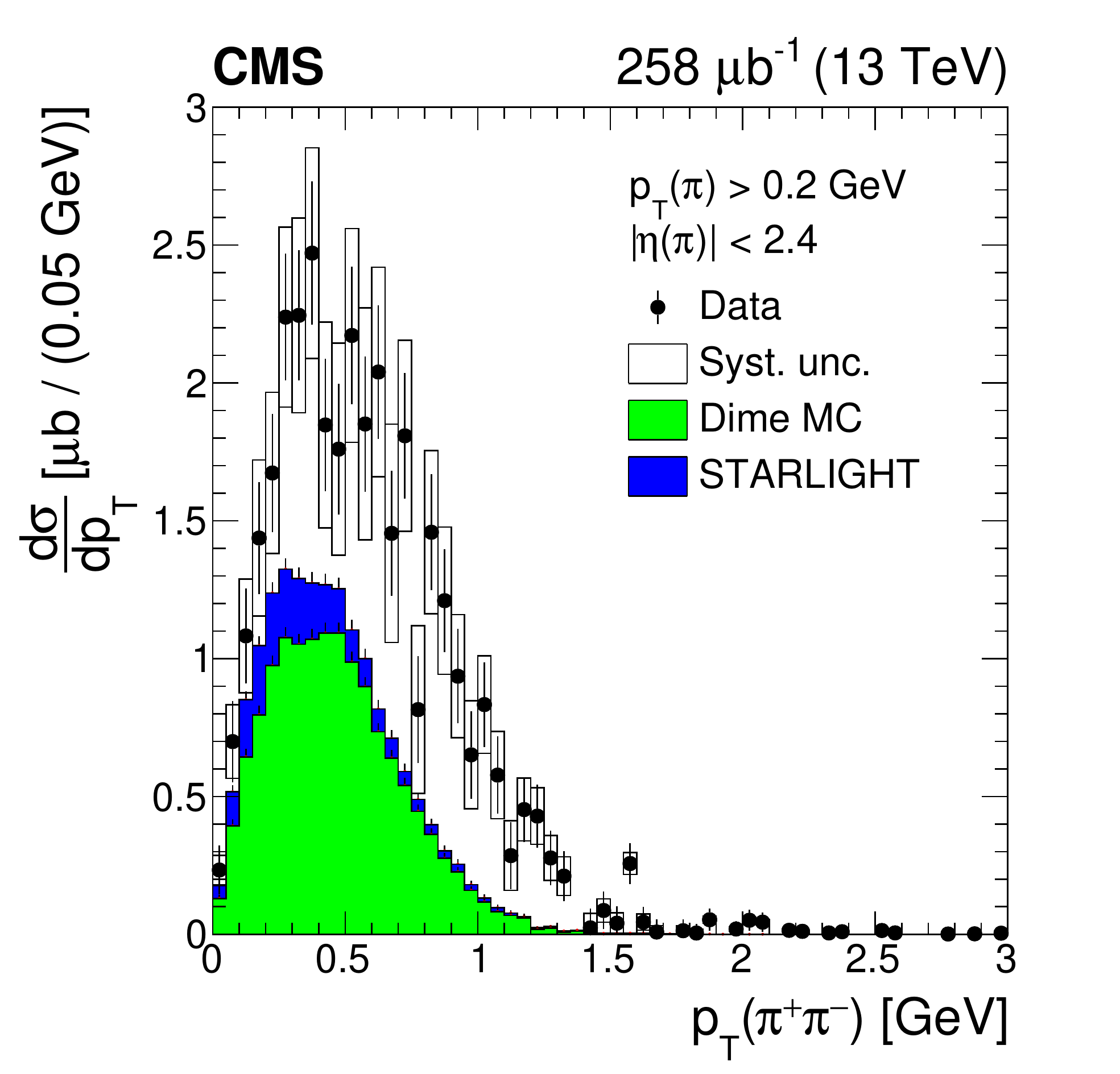} 
\includegraphics[width=0.45\textwidth]{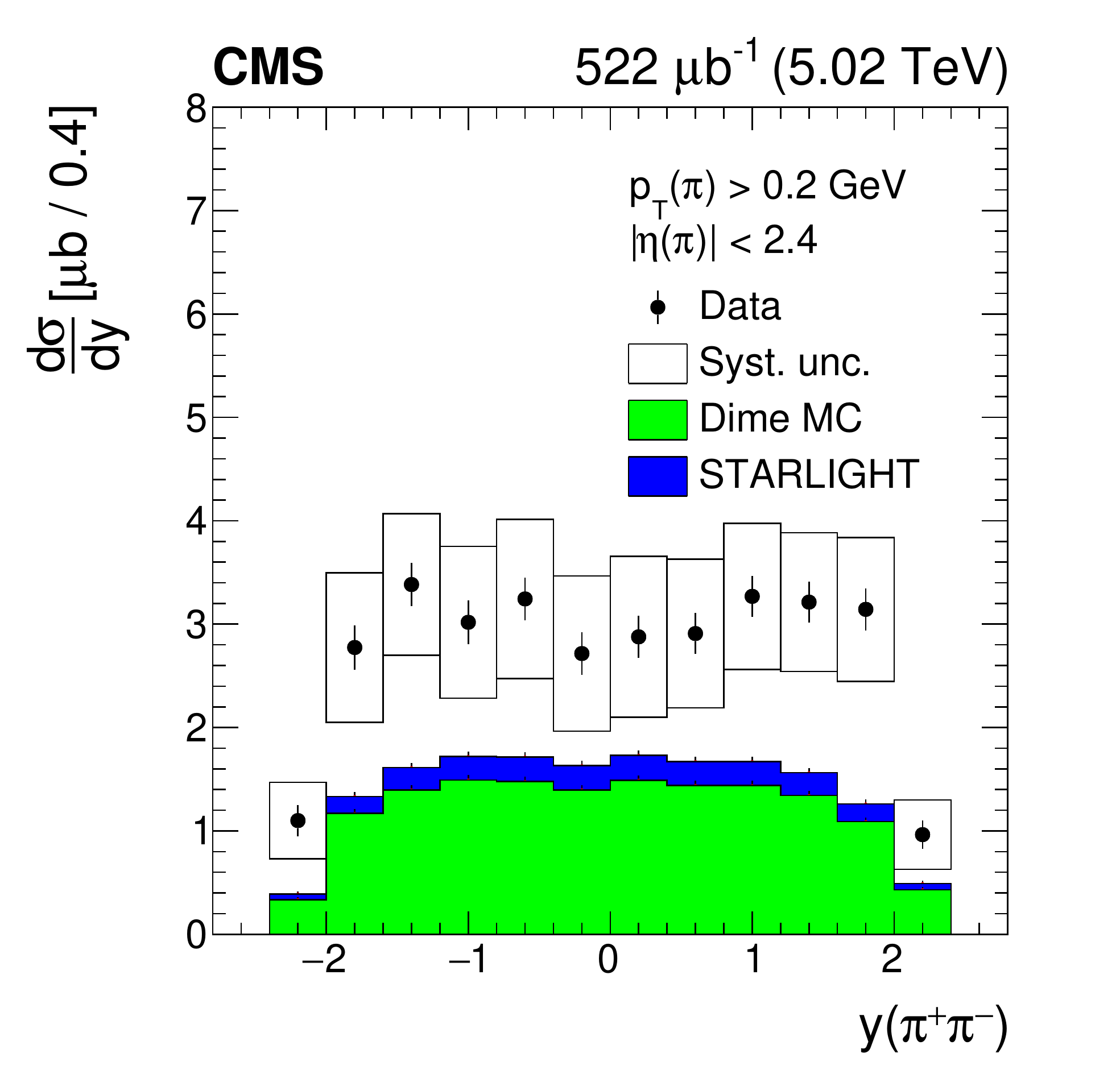} 
\includegraphics[width=0.45\textwidth]{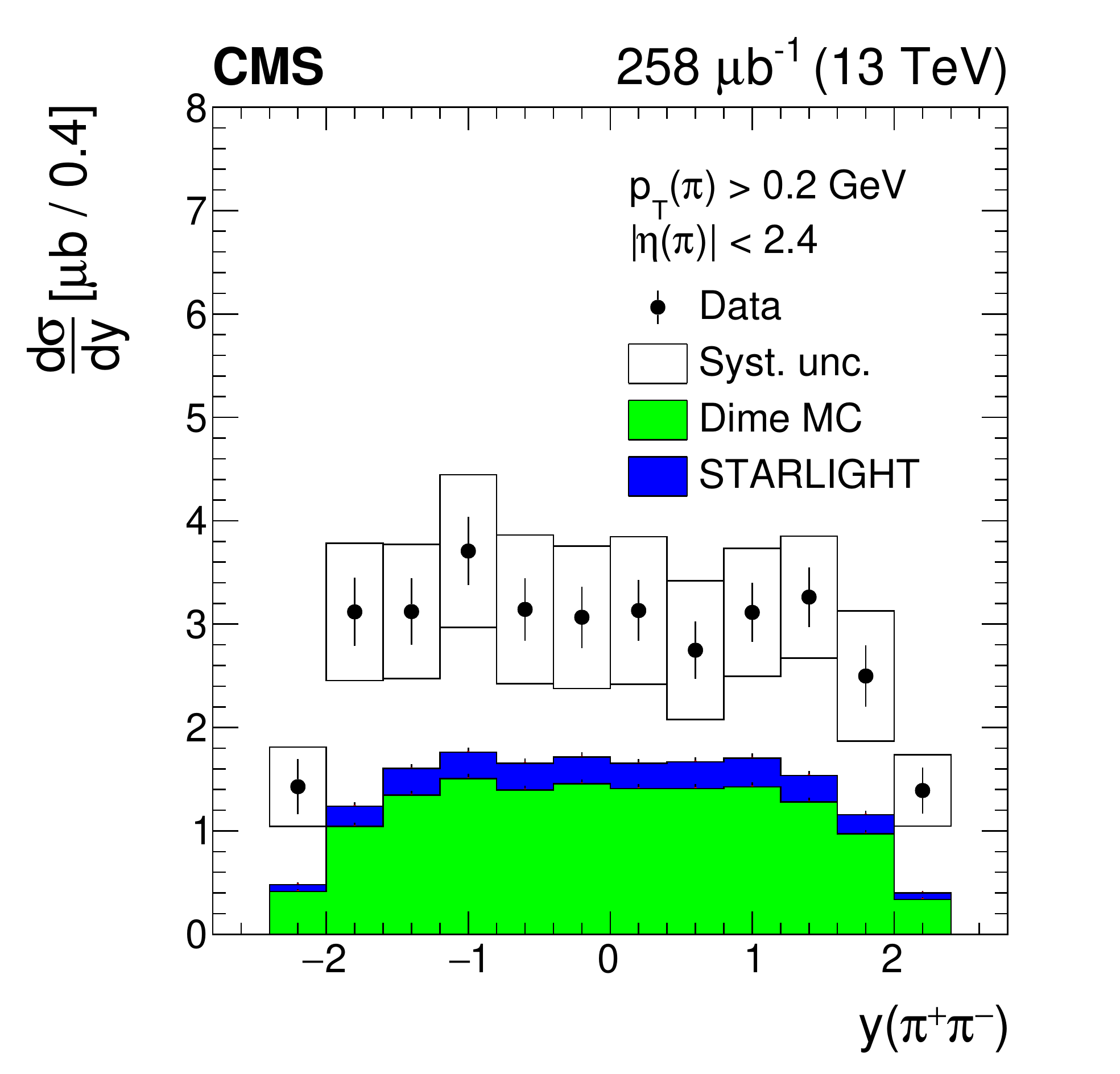} 
\caption{Differential cross sections as functions of mass (upper row), transverse momentum (middle row), and rapidity (bottom row), compared with generator-level simulations for the 5.02 (left) and 13\TeV (right) data sets. The error bars correspond to statistical, whereas the open boxes to systematic uncertainties.}
\label{fig:xsec}
\end{figure*}

The differential cross sections are calculated from the selected events as functions of the invariant mass, transverse momentum, and rapidity of the pion pair. These are shown in Fig.~\ref{fig:xsec} with the generator-level predictions from the \textsc{starlight} and \textsc{dime mc} generators, normalized to their cross sections. The MC generators provide an incomplete description of the available data, since they do not model the \PfDzP{500}, \PfDzP{980}, and \PfDtP{1270} resonances as mentioned in Section \ref{section:mc}.

{\tolerance=800 There is a peak at 800\MeV, which corresponds to the $\PGrzP{770}$ resonance. Since its quantum numbers $I^G(J^{PC}) = 1^+(1^{--})$ are forbidden in DPE processes, the $\PGrz$ mesons must be produced in VMP processes. The sharp drop visible around 1000\MeV is expected from previous measurements  \cite{Barberis:1999an,article:cdf} and can be attributed to the quantum mechanical interference of \PfDzP{980} with the continuum contribution. There is a prominent peak at 1200--1300\MeV, which corresponds to the \PfDtP{1270} resonance with $I^G(J^{PC}) = 0^+(2^{++})$ quantum numbers. This resonance is produced via a DPE process. \par}
 
Both \textsc{dime mc} and \textsc{starlight} underestimate the measured spectrum as these MC event generators do not model the forward dissociation of protons.
 
\begin{figure*}[htb]
\centering
\includegraphics[width=0.45\textwidth]{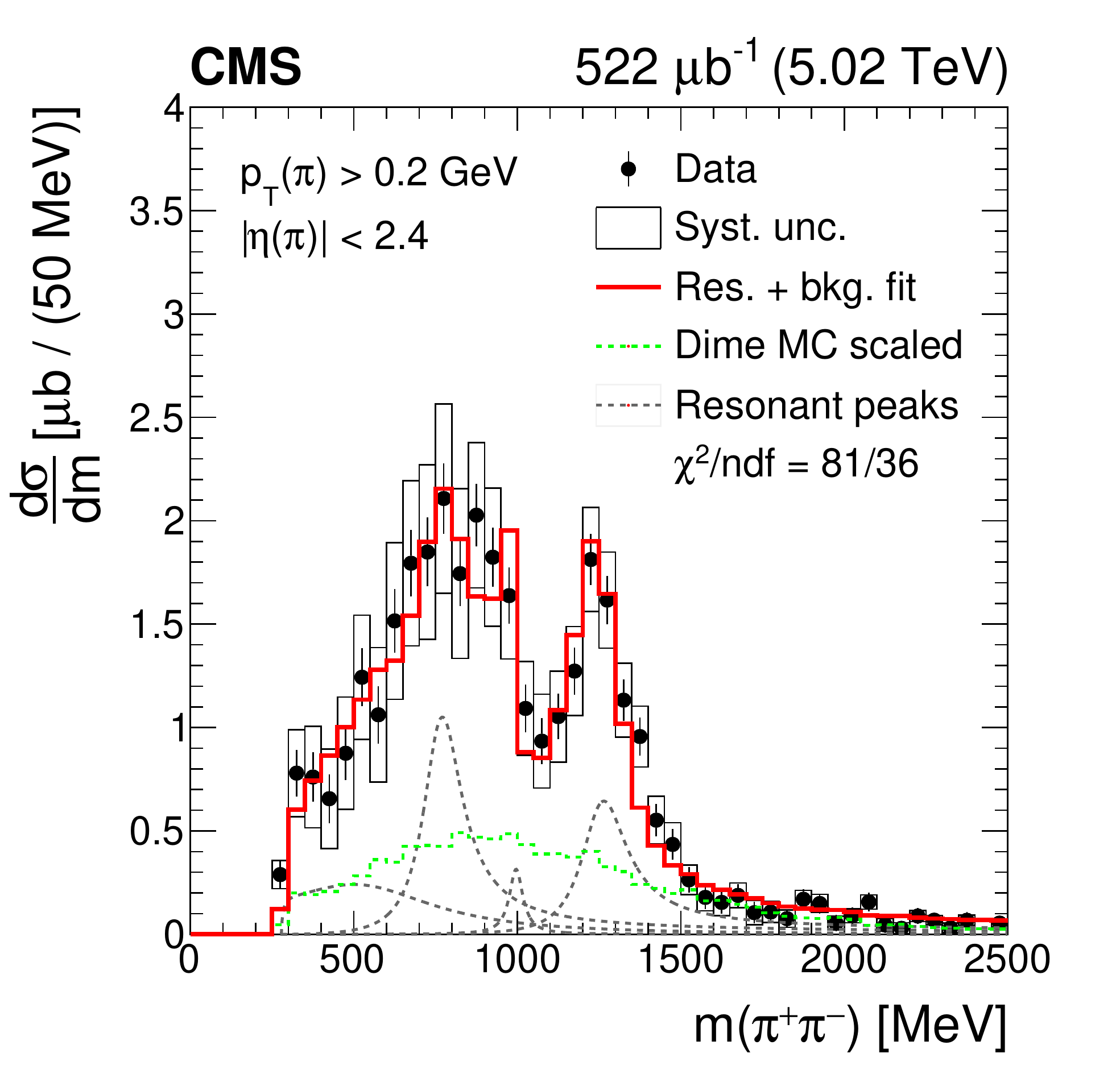} 
\includegraphics[width=0.45\textwidth]{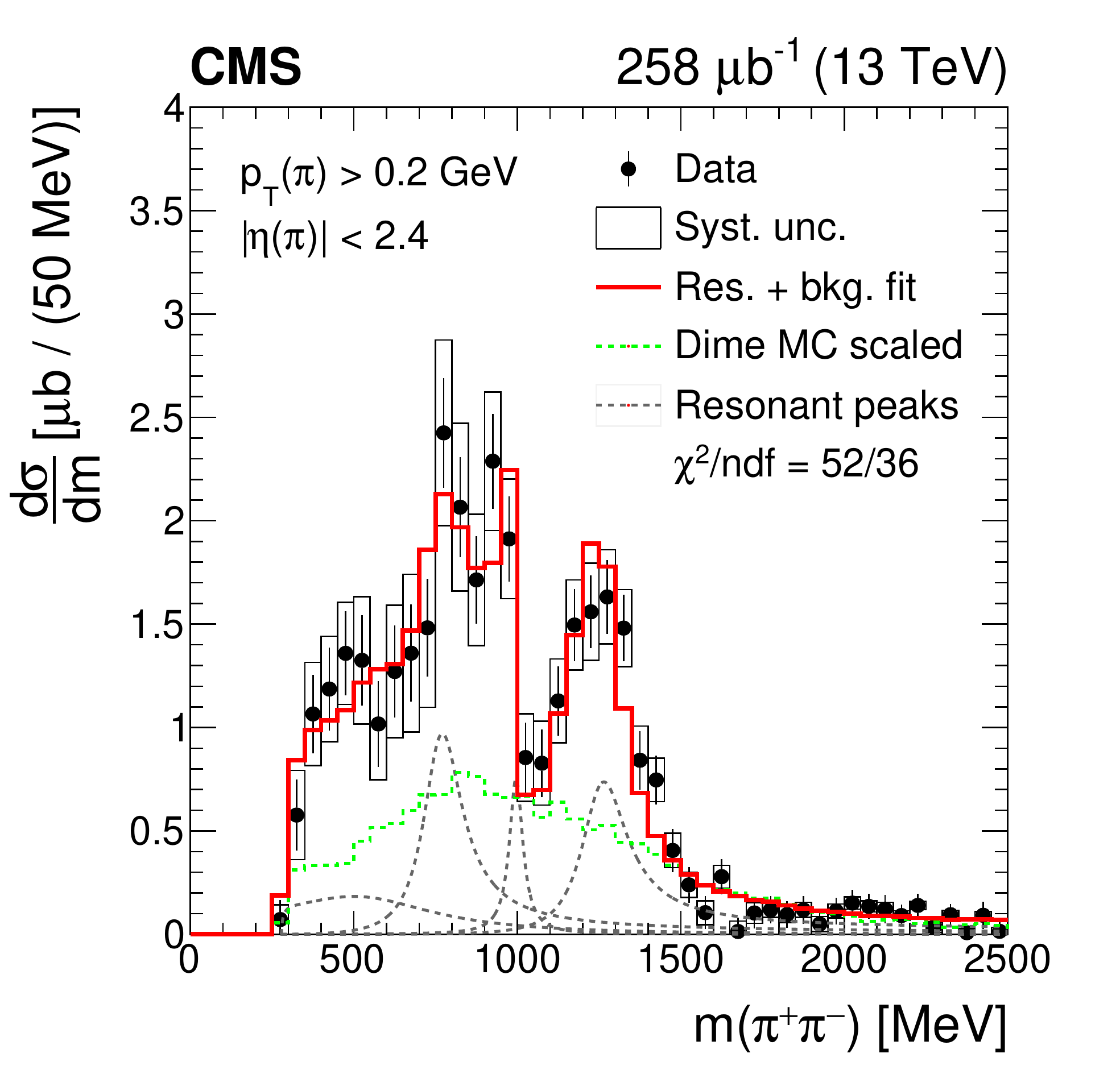} 
\caption{Fit to the measured cross section with the sum of four interfering relativistic Breit--Wigner functions convolved with a normal distribution (to account for the the experimental resolution of the detector) for the 5.02 (left) and 13\TeV (right) data sets. The error bars correspond to statistical, whereas the open boxes correspond to systematic uncertainties.}
\label{fig:fit}
\end{figure*}

The total cross section of the CEP process with two pions in the final state in the kinematic region $\pt(\PGp) > 0.2\GeV$ and $\abs{\eta(\PGp)} < 2.4$ is obtained by integrating the observed spectra in this region:
\begin{linenomath}
  \ifthenelse{\boolean{cms@external}}
{
\begin{multline}
\sigma_{\Pp\Pp\to \Pp'\Pp'\PGpp\PGpm} (\sqrt{s} = 5.02\TeV) = 32.6	\pm	0.7	\stat\\
\pm	6.0	\syst \pm	0.8	\lum
\mub,
\end{multline}
\begin{multline}
\sigma_{\Pp\Pp\to \Pp'\Pp'\PGpp\PGpm} (\sqrt{s} = 13\TeV) = 33.7	\pm	1.0	\stat\\
 \pm	6.2	\syst \pm	0.8	\lum\mub.
\end{multline}
}{
  \begin{align}
    \sigma_{\Pp\Pp\to \Pp'\Pp'\PGpp\PGpm} (\sqrt{s} = 5.02\TeV) &= 32.6	\pm	0.7	\stat \pm	6.0	\syst \pm	0.8	\lum\mub, \\
    \sigma_{\Pp\Pp\to \Pp'\Pp'\PGpp\PGpm} (\sqrt{s} = 13\TeV) &= 33.7	\pm	1.0	\stat \pm	6.2	\syst \pm	0.8	\lum\mub.
    \end{align}
}
\end{linenomath}

Below, it is demonstrated that the measured invariant $\PGpp\PGpm$ mass spectrum is well-described by the sum of the continuum distributions obtained from the \textsc{dime mc} model and four dominant resonances, modeled here by Breit-Wigner functions. In the fitting procedure the quantum mechanical interference effect and the detector resolution are also included.

The following fit function is used:
\begin{linenomath}
\ifthenelse{\boolean{cms@external}}
{
\begin{multline}
f(m) = \int G(m-m';\sigma) \Big[ \abs{A_\text{RBW}^{\PGrz}(m')}^2  \\ + \abs{A_\text{RBW}^{\PfDzP{500}}(m')\re^{\mathrm{i}\phi^{\PfDzP{500}}m'} + A_\mathrm{RBW}^{\PfDzP{980}}(m')\re^{\mathrm{i}\phi^{\PfDzP{980}}m'}  \\
+ A_\text{RBW}^{\mathrm{f}_2}(m')\re^{\mathrm{i}\phi^{\mathrm{f}_2}m'} + b \, B^{\textsc{dime}}(m')}^2 \Big] \rd m'.
\end{multline}
}
{
\begin{multline}
f(m) = \int G(m-m';\sigma) \Big[ \abs{A_\text{RBW}^{\PGrz}(m')}^2 + \abs{A_\text{RBW}^{\PfDzP{500}}(m')\re^{\mathrm{i}\phi^{\PfDzP{500}}m'} + A_\mathrm{RBW}^{\PfDzP{980}}(m')\re^{\mathrm{i}\phi^{\PfDzP{980}}m'}\\ 
+ A_\text{RBW}^{\mathrm{f}_2}(m')\re^{\mathrm{i}\phi^{\mathrm{f}_2}m'} + b \, B^{\textsc{dime}}(m')}^2 \Big] \rd m'.
\end{multline}
}
\end{linenomath}
Here $G(m;\sigma)$ is a Gaussian distribution with variance $\sigma$ and zero mean, $B^{\textsc{dime}}(m)$ is the nonresonant background estimated from the \textsc{dime mc} using the Orear-type form factor, and $b$ is a scale factor for the continuum contribution, and $\phi^{\PfDzP{500}}$, $\phi^{\PfDzP{980}}$, and $\phi^{\mathrm{f}_2}$ are phases that characterize interference effects. The $A_\text{RBW}^i(m)$ is the relativistic Breit--Wigner amplitude, which can be written as \cite{Jackson1964}:
\begin{linenomath}
\begin{align}
A_\text{RBW}^{i,J}(m) &= A_i \frac{\sqrt{mM_i\Gamma(m)}}{m^2-M_i^2 + i M_i \Gamma(m)}, \\
\Gamma(m) &= \Gamma_i \frac{M_i}{m}  \left[\frac{m^2 - 4m_{\PGp}^2}{M_i^2 - 4m_{\PGp}^2} \right]^{\frac{2J+1}{2}},
\end{align}
\end{linenomath}
where $A_i$, $M_i$, and $\Gamma_i$ are the yield, mass, and width of the resonance, respectively, $m_{\PGp}$ is the mass of charged pions, and $J$ is the total angular momentum of the resonance. According to Ref.\ \cite{Lebiedowicz:2016ioh}, the magnitude of the interference between the DPE and VMP processes is around 1\%, therefore no interference term is used between $\PGrz$ and DPE resonances. The convolution with the Gaussian distribution models the mass resolution of the detector.

The mass resolution ($\sigma$) is calculated by fitting the distribution of the difference between generator-level and reconstructed mass from the \textsc{starlight} and \textsc{dime mc} simulations. Based on these calculations, the mass resolution is found to vary from 9 to 14\MeV in the mass range 500--2000\MeV. In the final fit, an effective mass resolution of 11\MeV is used and the systematic uncertainty associated with this value is taken into account by repeating the fit with a mass resolution varying from 9 to 14\MeV. The resulting systematic uncertainty is 7--8\% for the yield of \PfDzP{980} and around 1--2\% for the yields of the \PfDzP{500}, \PGrzP{770}, and \PfDtP{1270} resonances.
The impact of the uncertainty in the multiparticle (exclusive $\PKp\PKm$ and $\Pp\PAp$) background yield is included by varying the background normalization in the fit by $\pm 14\%$ ($\pm 72\%$).
 
\begin{table*}[th]
\centering
\topcaption{Cross sections of the resonant processes in the $\pt(\PGp) > 0.2\GeV, \abs{\eta(\PGp)} < 2.4$ fiducial region, extracted from the simple model fit using the sum of the continuum distribution obtained from the \textsc{dime mc} model and four dominant resonances. The luminosity-related uncertainties are included in the systematic uncertainties. The \textsc{starlight} predictions for $\Pp\Pp \to \Pp'\Pp' \PGrz \to \Pp'\Pp' \PGpp\PGpm$ processes are 2.3 and 3.0\mub for 5.02 and 13\TeV, respectively, which is compatible with the fit results.}
\begin{tabular}{lrr}
\hline
\multicolumn{1}{c}{Resonance} & \multicolumn{2}{c}{$\sigma_{\Pp\Pp \to \Pp'\Pp' X \to \Pp'\Pp' \PGpp\PGpm} [\mu \mathrm{b}]$} \\
\hline
& \multicolumn{1}{c}{$\sqrt{s} = 5.02\TeV$} & \multicolumn{1}{c}{$\sqrt{s} = 13\TeV$} \\
\PfDzP{500}	& $	2.8	\pm		1.4			\stat\pm		2.2			\syst			$ & $	2.2	\pm		0.8			\stat\pm		1.3			\syst	$ \\
$\PGrzP{770}$	& $	4.7	\pm		0.9			\stat\pm		1.3			\syst			$ & $	4.3	\pm		1.3			\stat\pm		1.5			\syst	$ \\
\PfDzP{980}	& $	0.5	\pm		0.1			\stat\pm		0.1			\syst			$ & $	1.1	\pm		0.4			\stat\pm		0.3			\syst	$ \\
\PfDtP{1270}	& $	3.6	\pm		0.6			\stat\pm		0.7			\syst			$ & $	4.2	\pm		0.9			\stat\pm		0.8			\syst	$ \\
\hline
\end{tabular}
\label{tab:res}
\end{table*}

The masses and widths of \PGrzP{770} and \PfDtP{1270} resonances are fixed to the values of Ref.~\cite{book:pdg}. The mass and width of \PfDzP{500} and \PfDzP{980} are fixed according to the results from the most advanced calculations using dispersion relations \cite{GarciaMartin:2011jx}.

The fits are also performed with the mass and width of \PfDzP{500} and \PfDzP{980} varied according to their uncertainties \cite{book:pdg} and the resulting variation in the cross section of the resonances is added in quadrature to the other systematic uncertainty contributions. Furthermore the fit is repeated with the two other \textsc{dime mc} settings and the variation in the cross section is taken as an additional systematic uncertainty and added in quadrature to the other uncertainties.

The above simple model fit also provides values for the cross sections of the resonances; these are obtained by integrating the fitted squared amplitudes from the dipion threshold ($2m_{\PGp}$) to $M_i+5\Gamma_i$:
\begin{linenomath}
\begin{equation}
\sigma^\text{res}_i = \int_{2m_{\PGp}}^{M_i+5\Gamma_i} \abs{A^2_{\text{RBW},i}(m)} \rd m.
\end{equation}
\end{linenomath}
The fits are shown in Fig.~\ref{fig:fit} and the cross sections are summarized in Table \ref{tab:res}.

The model of interfering Breit--Wigner resonances with a continuum gives a good description of the data in the region of resonant peaks (below 1500\MeV). The cross sections for $\PGrzP{770}$ production calculated from the fits are slightly larger than the predicted values from \textsc{starlight}, which are 2.3 and 3.0\mub for 5.02 and 13\TeV, respectively. The differences can be attributed to the additional semiexclusive contribution that is not modeled by \textsc{starlight}. The values of the scale parameter $b$ are $0.7 \pm 0.2$ for 5.02\TeV and $1.1 \pm 0.3$ for 13\TeV, and therefore they are consistent within uncertainties for the two energies.

\section{Summary}
The cross sections for central exclusive pion pair production have been measured in $\Pp\Pp$ collisions at 5.02 and 13\TeV center-of-mass energies. Exclusive events are selected by vetoing additional energy deposits in the calorimeters and by requiring two oppositely charged pions identified via their mean energy loss in the tracker detectors. These events are used together with correction factors to obtain invariant mass, transverse momentum, and rapidity distributions of the $\PGpp\PGpm$ system. The measured total exclusive $\PGpp\PGpm$ production cross section is $32.6	\pm	0.7	\stat \pm	6.0	\syst \pm	0.8\lum$ and $33.7	\pm	1.0	\stat \pm	6.2	\syst \pm	0.8\lum$\mub for 5.02 and 13\TeV, respectively. The observed mass spectrum exhibits resonant structures, which can be fitted with a simple model containing four interfering Breit-Wigner functions, corresponding to the \PfDzP{500}, $\PGrzP{770}$, \PfDzP{980}, and \PfDtP{1270} resonances, and a continuum contribution modeled by the \textsc{dime mc}. The exclusive production cross sections are extracted from this fit.
The obtained cross sections of $\PGrzP{770}$ production are higher than the \textsc{starlight} model prediction, which can be explained by the presence of semiexclusive production which is not modeled by the \textsc{starlight} generator.
\ifthenelse{\boolean{cms@external}}{\clearpage}{}
\begin{acknowledgments}
  We congratulate our colleagues in the CERN accelerator departments for the excellent performance of the LHC and thank the technical and administrative staffs at CERN and at other CMS institutes for their contributions to the success of the CMS effort. In addition, we gratefully acknowledge the computing centers and personnel of the Worldwide LHC Computing Grid for delivering so effectively the computing infrastructure essential to our analyses. Finally, we acknowledge the enduring support for the construction and operation of the LHC and the CMS detector provided by the following funding agencies: BMBWF and FWF (Austria); FNRS and FWO (Belgium); CNPq, CAPES, FAPERJ, FAPERGS, and FAPESP (Brazil); MES (Bulgaria); CERN; CAS, MoST, and NSFC (China); COLCIENCIAS (Colombia); MSES and CSF (Croatia); RPF (Cyprus); SENESCYT (Ecuador); MoER, ERC IUT, PUT and ERDF (Estonia); Academy of Finland, MEC, and HIP (Finland); CEA and CNRS/IN2P3 (France); BMBF, DFG, and HGF (Germany); GSRT (Greece); NKFIA (Hungary); DAE and DST (India); IPM (Iran); SFI (Ireland); INFN (Italy); MSIP and NRF (Republic of Korea); MES (Latvia); LAS (Lithuania); MOE and UM (Malaysia); BUAP, CINVESTAV, CONACYT, LNS, SEP, and UASLP-FAI (Mexico); MOS (Montenegro); MBIE (New Zealand); PAEC (Pakistan); MSHE and NSC (Poland); FCT (Portugal); JINR (Dubna); MON, RosAtom, RAS, RFBR, and NRC KI (Russia); MESTD (Serbia); SEIDI, CPAN, PCTI, and FEDER (Spain); MOSTR (Sri Lanka); Swiss Funding Agencies (Switzerland); MST (Taipei); ThEPCenter, IPST, STAR, and NSTDA (Thailand); TUBITAK and TAEK (Turkey); NASU (Ukraine); STFC (United Kingdom); DOE and NSF (USA). 
 
 \hyphenation{Rachada-pisek} Individuals have received support from the Marie-Curie program and the European Research Council and Horizon 2020 Grant, contract Nos.\ 675440, 752730, and 765710 (European Union); the Leventis Foundation; the A.P.\ Sloan Foundation; the Alexander von Humboldt Foundation; the Belgian Federal Science Policy Office; the Fonds pour la Formation \`a la Recherche dans l'Industrie et dans l'Agriculture (FRIA-Belgium); the Agentschap voor Innovatie door Wetenschap en Technologie (IWT-Belgium); the F.R.S.-FNRS and FWO (Belgium) under the ``Excellence of Science -- EOS" -- be.h project n.\ 30820817; the Beijing Municipal Science \& Technology Commission, No. Z191100007219010; the Ministry of Education, Youth and Sports (MEYS) of the Czech Republic; the Deutsche Forschungsgemeinschaft (DFG) under Germany's Excellence Strategy -- EXC 2121 ``Quantum Universe" -- 390833306; the Lend\"ulet (``Momentum") Program and the J\'anos Bolyai Research Scholarship of the Hungarian Academy of Sciences, the New National Excellence Program \'UNKP, the NKFIA research grants 123842, 123959, 124845, 124850, 125105, 128713, 128786, and 129058 (Hungary); the Council of Science and Industrial Research, India; the HOMING PLUS program of the Foundation for Polish Science, cofinanced from European Union, Regional Development Fund, the Mobility Plus program of the Ministry of Science and Higher Education, the National Science Center (Poland), contracts Harmonia 2014/14/M/ST2/00428, Opus 2014/13/B/ST2/02543, 2014/15/B/ST2/03998, and 2015/19/B/ST2/02861, Sonata-bis 2012/07/E/ST2/01406; the National Priorities Research Program by Qatar National Research Fund; the Ministry of Science and Education, grant no. 14.W03.31.0026 (Russia); the Tomsk Polytechnic University Competitiveness Enhancement Program and ``Nauka" Project FSWW-2020-0008 (Russia); the Programa Estatal de Fomento de la Investigaci{\'o}n Cient{\'i}fica y T{\'e}cnica de Excelencia Mar\'{\i}a de Maeztu, grant MDM-2015-0509 and the Programa Severo Ochoa del Principado de Asturias; the Thalis and Aristeia programs cofinanced by EU-ESF and the Greek NSRF; the Rachadapisek Sompot Fund for Postdoctoral Fellowship, Chulalongkorn University and the Chulalongkorn Academic into Its 2nd Century Project Advancement Project (Thailand); the Kavli Foundation; the Nvidia Corporation; the SuperMicro Corporation; the Welch Foundation, contract C-1845; and the Weston Havens Foundation (USA). \end{acknowledgments}

\bibliography{auto_generated}
\cleardoublepage \appendix\section{The CMS Collaboration \label{app:collab}}\begin{sloppypar}\hyphenpenalty=5000\widowpenalty=500\clubpenalty=5000\input{FSQ-16-006-authorlist.tex}\end{sloppypar}
%%% END EDITABLE REGION %%%
\end{document}

%% file: FSQ-16-006-authorlist.tex
\vskip\cmsinstskip
\textbf{Yerevan Physics Institute, Yerevan, Armenia}\\*[0pt]
A.M.~Sirunyan$^{\textrm{\dag}}$, A.~Tumasyan
\vskip\cmsinstskip
\textbf{Institut f\"{u}r Hochenergiephysik, Wien, Austria}\\*[0pt]
W.~Adam, F.~Ambrogi, T.~Bergauer, J.~Brandstetter, M.~Dragicevic, J.~Er\"{o}, A.~Escalante~Del~Valle, M.~Flechl, R.~Fr\"{u}hwirth\cmsAuthorMark{1}, M.~Jeitler\cmsAuthorMark{1}, N.~Krammer, I.~Kr\"{a}tschmer, D.~Liko, T.~Madlener, I.~Mikulec, N.~Rad, J.~Schieck\cmsAuthorMark{1}, R.~Sch\"{o}fbeck, M.~Spanring, D.~Spitzbart, W.~Waltenberger, J.~Wittmann, C.-E.~Wulz\cmsAuthorMark{1}, M.~Zarucki
% ...
\vskip\cmsinstskip
\dag: Deceased\\
1:  Also at Vienna University of Technology, Vienna, Austria\\
% ...
